\documentclass[fleqn,usenatbib,onecolumn]{mnras}
\usepackage{newtxtext,newtxmath}
\usepackage[T1]{fontenc}

\DeclareRobustCommand{\VAN}[3]{#2}
\let\VANthebibliography\thebibliography
\def\thebibliography{\DeclareRobustCommand{\VAN}[3]{##3}\VANthebibliography}

\usepackage{graphicx}	% Including figure files
\usepackage{amsmath}	% Advanced maths commands
\usepackage{amssymb}	% Extra maths symbols
\usepackage{natbib}
\usepackage{comment}
\usepackage{threeparttable}
\def\pd{{\rm d}}

\def\rs{\rho_{\rm S}}

\title[Multivariate distribution with vine copula]{
  Constructing a multivariate distribution function with a vine copula:  
  toward multivariate luminosity and mass functions}

\author[T.\ T.\ Takeuchi \& K.\ T.\ Kono]
  {Tsutomu T.~Takeuchi$^{1,2}$\thanks{E-mail: takeuchi.tsutomu@g.mbox.nagoya-u.ac.jp} \& Kai T.\ Kono$^1$\\
  $^{1}$Division of Particle and Astrophysical Science, Nagoya University, Furo-cho, Chikusa-ku, Nagoya 464-8602, Japan\\
  $^{2}$The Research Center for Statistical Machine
  Learning, the Institute of Statistical Mathematics, 10--3 Midori-cho, Tachikawa, Tokyo 190--8562, Japan
}

\date{Accepted XXX. Received YYY; in original form ZZZ}

\pubyear{2015}

\begin{document}
\label{firstpage}
\maketitle

%macros
\newtheorem{theorem}{Theorem}
\newtheorem{definition}{Definition}

\maketitle

\begin{abstract}
The need for a method to construct multidimensional distribution function is increasing recently, in the era of huge multiwavelength surveys. 
We have proposed a systematic method to build a bivariate luminosity or mass function of galaxies by using a copula.  %\citep{takeuchi2010b}. 
It allows us to construct a distribution function when only its marginal distributions are known, and we have to estimate the dependence structure from data. 
A typical example is the situation that we have univariate luminosity functions at some wavelengths for a survey, but the joint distribution is unknown. 
Main limitation of the copula method is that it is not easy to extend a joint function to higher dimensions ($d > 2$), except some special cases like multidimensional Gaussian. 
Even if we find such a multivariate analytic function in some fortunate case, it would often be inflexible and impractical. 
In this work, we show a systematic method to extend the copula method to unlimitedly higher dimensions by a vine copula. 
This is based on the pair-copula decomposition of a general multivariate distribution. 
We show how the vine copula construction is flexible and extendable. 
We also present an example of the construction of a stellar mass--atomic gas--molecular gas 3-dimensional mass function. 
We demonstrate the maximum likelihood estimation of the best functional form for this function, as well as a proper model selection via vine copula. 
\end{abstract}


\section{Introduction}\label{sec:introduction}

Galaxies evolve in various aspects. 
Individual galaxies change their physical properties through merging of their host dark matter halos, merging of galaxies, star formation, chemical evolution, infall of matter from the large-scale structure, etc. 
This aspect of galaxy evolution is, say, a life history of galaxies. 
The collection of the individual life history of galaxies, combined with the evolving cosmological condition, drives the collective evolution of galaxies. 
To describe this "sociological" galaxy evolution in a statistical sense, the luminosity function (LF) and/or mass function (MF) of galaxies play a fundamental role (e.g. \citealt{binggeli88,
%lin96,
takeuchi00b,
blanton01,
delapparent03,
willmer06,
johnston2011,
moffett2016,
koprowski2017,
lake2017,
lopez_sanjuan2017,
wright2017,
bhatawdekar2019}. 
Even though the LF (MF) is a result of highly complicated and entangled physical processes, still it is the first statistic to be examined from observations. 

Now, studies of galaxy evolution is facing the time for drastic change by multiband large surveys. 
Indeed, all of the modern large surveys are performed at multiband. 
Connecting the LFs (MFs) obtanied at different wavelengths is expected to provide us with a new insight to the fundamental physics to drive galaxy evolution \citep[e.g.][and references therein]{mashian2016,vallini2016,caplar2018,dutta2020}.  
However, it is not easy to determine the corresponding multivariate 
function from its marginal distributions, if the distribution is not multivariate Gaussian.
As widely known, galaxy LFs are fairly well described by the Schechter function \citep{schechter76} (stellar and gas components) or double-power-law type function \citep[e.g.][]{saunders00,takeuchi03b} (dust, radio continuum and X-ray emission), both of which are far different from the Gaussian distribution. 
In such a case, there exist infinitely many distributions with the same marginals even if the correlation structure is specified.
In astronomical applications, a multivariate distribution has been constructed based on a primary-selection wavelength \citep[e.g.][]{mobasher93,choloniewski85,chapman03,schafer07,calette2018,rodriguez_puebla2020}.
A thorough and comprehensive discussion on this method is found in, for example, \citet{rodriguez_puebla2020}.  
Though these works are well designed in their own purposes, we often want to have a multivariate PDF estimation method without a specific primary selection in modern astrophysical analysis. 
%Further, analytic bivariate distribution models are often required to interpret the distributions obtained by nonparametric methods \citep[e.g.][]{cross02,ball06,driver06}.
Thus, a general method to construct a multivariate 
distribution function with pre-defined marginal distributions and dependence structure has long been desired.

Such a function has been commonly used to analyze two covariate random variables, particularly extensively in econometrics and mathematical finance.
This is the so-called ``copula''\footnote{In \citet{takeuchi2010b}, we used "copulas" as its plural form. 
However, since this terminology is a Latin feminine noun, we use "copulae" instead in this paper.}.
In a bivariate context, copulae are obviously useful to define nonparametric measures of dependence for pairs of random variables (e.g. \citealt{johnson77,trivedi05}; for a recent review, see \citealt{lin2014}).
In astrophysics, however, copulae started to attract researchers' attention relatively recently \citep[e.g.][]{benabed09,jiang09,koen09,scherrer10,takeuchi2010b}. 
After a decade since then, the copula method is getting gradually known to the astronomical community: now it is applied to bivariate luminosity function of galaxies  \citep[e.g.][]{takeuchi2013,andreani2014,gunawardhana2015,andreani2018,yuan2018}, 
completeness problems in galaxy surveys \citep[e.g.][]{johnston2012}, 
cosmology with gravitational lensing \citep[e.g.][]{sato2010,sato2011,lin2015,simon2017}, 
time series analysis of bivariate sequence \citep[e.g.][]{jo2019a} 
and many other astrophysical applications \citep[e.g.][]{jiang2015,koen2017,jo2019b}. 

When we have introduced the copula method to the galactic astrophysics and cosmology in \citet{takeuchi2010b}, practical application of the copula was restricted to the bivariate problems. 
This is because of the fact that the copula method was not easy to extend to higher dimensions ($d > 2$), except some special cases like Gaussian. 
Further, even if we find such a multivariate analytic function in some very fortunate case, it would probably be very inflexible and impractical, for example, to a realistic statistical estimation in galaxy surveys. 
Actually, however, a method to improve the copula method and resolve the difficulty to multivariate extension was introduced just some years before \citet{takeuchi2010b}. 
This is based on the decomposition of a general multivariate distribution: 
a multivariate probability density function can be factorized into a bivariate copulae and univariate density functions.
Since we have a rich theoretical method of bivariate copulae, this means that we can extend our methodology to any higher dimension problems \citep[e.g.][and references therein]{aas2009}. 
However, since this decomposition is not unique, we need to sort it out to have a systematic procedure. 
For this purpose, we introduce the concept of vine copula, invented in the field of graphical modeling \citep{bedford2002}. 
Since the work of \citet{aas2009}, vine copulae have been applied to vastly wide range of fields: 
financial risk management \citep[e.g.][]{sriboonchitta2014,aas2016,allen2017,nagler2019}, 
insurance \citep[e.g.][]{peters2014,mejdoub2018,shi2018}, 
weather forecast and engineering \citep[e.g.][]{alidoost2019,kloubert2020,torabi2020}, 
multivariate time series analysis \citep[e.g.][]{almeida2016,jager2017,acar2019}, 
spatio-temporal analysis \citep[e.g.][]{graler2011,graler2014,callau_poduje2018}, 
and technological applications \citep[e.g.][]{xu_d2017, xu_m2017,khuntia2019}, among many others.
 
In this work, we show a systematic method to extend the copula to unlimitedly higher dimensions by a vine copula method\footnote{After the submission of this manuscript, a similar work by \citet{vio2020} appeared on arXiv. 
Readers are also guided to their article as a different astrophysical application of the vine copula. 
}. 

This paper is organized as follows: in Section~\ref{sec:formulation} we briefly review the basics of copula. 
Then we introduce the central concept of this work, vines, and formulate the systematic construction of a multivariate copula with vines. 
In Section~\ref{sec:mlf}, we make use of these copulae to construct a MMF of
galaxies. 
We discuss some implications and further applications in Section~\ref{sec:mlf}.
Section~\ref{sec:conclusion} is devoted to summary and conclusions.

Throughout this paper, we adopt a cosmological model $(h, \Omega_{\rm M0}, \Omega_{\rm \Lambda0})
= (0.7, 0.3, 0.7)$ ($h \equiv H_0/100 [\mbox{km}\,\mbox{s}^{-1}] \, \mbox{Mpc}^{-1}$).

\section{Formulation}
\label{sec:formulation}

\subsection{Copula}
\label{subsec:copula}

First we briefly review the concept of copula. 
In short, copulae are functions that relate joint multivariate distribution functions (DFs) to their one-dimensional marginal DFs\footnote{As in T10\nocite{takeuchi2010b}, the DF stands for a cumulative distribution function in statistical terminology. 
To avoid confusion, we use a term "probability density function (PDF)" to refer to a distribution function commonly used in physics.
In this paper (and statistical literature in general), we distinguish a DF and PDF by an upper and lower case, respectively (e.g. $F(x)$ stands for a certain DF, and $f(x)$ is its PDF).}.
Using a copula $C$, any multivariate DF, $G$, can be expressed with margins $F_1, F_2, \dots, F_d$ as
\begin{eqnarray}
  G(x_1,\dots,x_d) = C[F(x_1),\dots, F_d(x_d)] \;. \label{eq:copula}
\end{eqnarray}
This is guaranteed by Sklar's theorem \citep{sklar59}.
Especially, if $F_1, \dots, F_d$ are continuous, then $C$ is unique.
A comprehensive proof of Sklar's theorem is found in e.g. \citet{nelsen06}.
This theorem gives a basis that {\it any} multivariate DF with given margins is expressed with a form of eq.~(\ref{eq:copula}).
If we want a more familiar form, a PDF of $G(x_1, \dots, x_d)$, $g(x_1, \dots, x_d)$, is written as
\begin{eqnarray}\label{eq:copula_density}
  g(x_1,\dots,x_n) = \frac{\partial^d C[F_1(x_1),\dots, F_d(x_d)]}{\partial x_1 \dots \partial x_d} f_1(x_1) \dots f_d(x_d)
    \equiv c[F_1(x_1), \dots, F_d(x_d)] f_1(x_1) \dots f_d(x_d)
\end{eqnarray}
where $f_1(x_1), \dots, f_d(x_d)$ are PDFs of $F_1(x_1), \dots, F_d(x_d)$, respectively.
Here, a function $c[F_1(x_1), \dots, F_d(x_d)]$ is referred to as the copula density of $C$. 
For more detailed (but not too rigorous) definitions, readers are guided to \citet[][hereafter T10]{takeuchi2010b}. 

The most important statistical aspect of bivariate DFs is their
dependence properties between variables.
Since the dependence can never be given by the marginals of a DF, this is the most
nontrivial information which a bivariate DF provides.
Since any bivariate DFs are described by Equation~(\ref{eq:copula}), all the information
on the dependence is carried by their copulae.

For practical data analysis, a measure of dependence is useful for the interpretation of a result.
The Pearson's product-moment correlation coefficient $\rho$ is the most frequently used dependence measure  for physical scientists (and others). 
For a while in this paragraph, we focus on the bivariate PDF since we consider correlation measures. 
The bivariate PDF of $x_1$ and $x_2$, $g(x_1,x_2)$, is written as
\begin{eqnarray}\label{eq:copula_density_2dim}
  g(x_1,x_2) = \frac{\partial^2 C[F_1(x_1),F_2(x_2)]}{\partial x_1 \partial x_2} f_1(x_1) f_2(x_2)
    = c[F_1(x_1),F_2(x_2)] f_1(x_1) f_2(x_2). 
\end{eqnarray}
Then the correlation coefficient $\rho$ is expressed as
\begin{eqnarray}\label{eq:corr}
  \rho = \frac{\int (x_1 - \bar{x_1})(x_2 - \bar{x_2}) g(x_1,x_2) 
    \pd x_1 \pd x_2}{\sqrt{\int (x_1 - \bar{x_1})^2 f_1(x_1) \pd x_1 
    \int (x_2 - \bar{x_2})^2 f_2(x_2) \pd x_2}}  = 
  \frac{\int (x_1 - \bar{x_1})(x_2 - \bar{x_2})  c[(F_1(x_1),F_2(x_2)] f_1(x_1) f_2(x_2) \pd x_1 \pd x_2}{
    \sqrt{\int (x_1 - \bar{x_1})^2 f_1(x_1) \pd x_1 \int (x_2 - \bar{x_2})^2 f_2(x_2) \pd x_2}} \;.
\end{eqnarray}
We observe that Equation~(\ref{eq:corr}) depends not only on the dependence of two variables (copula part) but also its marginals $f_1(x_1), f_2(x_2)$, i.e.,
{\it the linear correlation coefficient $\rho$ does not measure the dependence purely.}
Then, sometimes a genuine measure of dependence, e.g. Spearman's $\rs$ or Kendall's $\tau$ would be more appropriate.
Spearman's rank correlation is a nonparametric version of Pearson's correlation using a rank of data.
The population version of Spearman's $\rs$ is expressed by copula as
\begin{eqnarray}\label{eq:spearman_copula}
  \rs = 12 \int_0^1 \int_0^1 u_1 u_2 \pd C(u_1,u_2) - 3 
    = 12 \int_0^1 \int_0^1 C(u_1, u_2) \pd u_1 \pd u_2 -3 \;. 
\end{eqnarray}
Kendall's $\tau$ is also expressed in a simple form in terms of copula as
\begin{eqnarray}\label{eq:kendall_copula}
    \tau =
    4 \int_0^1 \int_0^1 C(u_1,u_2) \pd C(u_1,u_2) - 1 =
    4 \int_0^1 \int_0^1 C(u_1, u_2) c(u_1,u_2) \pd u_1 \pd u_2 -1 \;. 
\end{eqnarray}
The derivation of these equations are found in T10\nocite{takeuchi2010b}. 
In Equations~(\ref{eq:spearman_copula}) and (\ref{eq:kendall_copula}) are {\it independent of the distributions $F_1, F_2$, or $G$} and depend only on the dependence structure, i.e., a copula. 
This is the reason why the two dependence measures are almost 
always used in the context of copulae in the literature.

What we have from surveys are usually multivariate datasets, and we do not know the functional form of a multivariate DF from which the data are sampled.
Namely, there is infinite degrees of freedom for a set of copulae to choose. 
For a bivariate case, it might be still possible to restrict a class of functions for a copula \citep[e.g.][]{takeuchi2013,andreani2014,gunawardhana2015,andreani2018}.
However, it would be almost impossible to have an intuition to choose an appropriate family of a single multivariate copulae/copula densities for a certain survey data. 
To make the problem practically more accessible, we need a systematic construction method of a multivariate copula from a lower-dimensional information. 
We introduce such a method in the following. 

\subsection{Vine copula}
\label{subsec:vine_copula}

Here we introduce a vine copula as a systematic method to factorize a multivariate PDF as above. 
A vine is a concept originally introduced in the field of 
graphical modeling \citep{bedford2002}. 

\subsubsection{Factorization of a PDF}

As we mentioned in \ref{sec:introduction}, this is based on the decomposition of a general multivariate distribution. 
Let $f(x_1, \dots, x_d)$ be a joint PDF of a set of $d$-dimensional vector stochastic variable $\vec{X} = (X_1, \dots, X_d)$. 
First, recall the formula of conditional probability
\begin{eqnarray}
  f(A,B) = f(B|A)f(A) \; ,
\end{eqnarray}
where $A$ and $B$ are events.
If we apply this formula to the above PDF, we have 
\begin{eqnarray}
  f(x_1, \dots, x_d) &=& f_{2\dots d|1}(x_2, \dots, x_d|x_1)f_{1}(x_1) \nonumber \\
  &=& f_{3\dots d|12}(x_3, \dots, x_d|x_1, x_2) f_{2|1}(x_2|x_1)f_{1}(x_1) \nonumber \\
%  &=& f_{4\dots d|123}(x_4, \dots, x_d|x_1, x_2, x_3) f_{3|12}(x_3|x_1, x_2) f_{2|1}(f_2|x_1)f_{1}(x_1) \nonumber \\
  &\vdots& \nonumber \\
  &=& f_{d|123\dots d-1}(x_d|x_1, \dots, x_{d-1}) \cdots f_{2|1}(x_2|x_1)f_{1}(x_1) 
\end{eqnarray}
which is known as the chain rule. 
We start from this well-known mathematical formula. 
By using a bivariate copula density
\begin{eqnarray}
  f_{12}(x_1, x_2) = c\left[F_1(x_1), F_2(x_2) \right]f_1(x_1)f_2(x_2) \;, 
\end{eqnarray}
the conditional probability can be expressed as
\begin{eqnarray}
  f_{2|1}(x_2|x_1) &=& \frac{f_{12}(x_1,x_2)}{f_1(x_1)} \nonumber \\
  &=& c_{12}\left[F_1(x_1), F_2(x_2) \right]f_2(x_2) \; .
\end{eqnarray}
Similarly, for $f_{23|1}$, 
\begin{eqnarray}
  f_{23|1}(x_2,x_3|x_1) 
  &=& c_{23|1}\left[F_{2|1}(x_2|x_1), F_{3|1}(x_3|x_1) \right]f_{2|1}(x_2|x_1)f_{3|1}(x_3|x_1) \; .
\end{eqnarray}
Since 
\begin{eqnarray}
  \frac{f_{23|1}(x_2, x_3|x_1)}{f_{2|1}(x_2|x_1)} = f_{3|12}(x_3|x_1, x_2) \; , 
\end{eqnarray}
we obtain
\begin{eqnarray}
  f_{3|12}(x_3|x_1,x_2) &=& c_{23|1}\left[F_{2|1}(x_2|x_1), F_{3|1}(x_3|x_1) \right]f_{3|1}(x_3|x_1) \nonumber \\
  &=& c_{23|1}\left[F_{2|1}(x_2|x_1), F_{3|1}(x_3|x_1) \right]c_{13}\left[F_1(x_1), F_3(x_1)\right]f_3(x_3) \; . 
\end{eqnarray}
We can generalize the formula.  
Set $\vec{v}\equiv (v_1, \dots, v_k) = x_{i_1}, \dots, x_{i_k}$. 
If we define $\vec{v}_{-j} \equiv (v_1, \dots, v_{j-1}, v_{j+1}, \dots, v_k)$, the following formula holds
\begin{eqnarray}\label{eq:copula_density_coef}
  f(x|\vec{v}) &=& c_{x\vec{v}_j|\vec{v}_{-j}} \left[F(x)|\vec{v}_{-j}), F_{v_j}(v_j|\vec{v}_{-j}) \right]f(x|\vec{v}_{-j})\;. 
\end{eqnarray}
This is a purely mathematical, direct result of the formula of conditional probability. 

We should note that such a decomposition is {\sl not unique} if we consider a permutation of the labels of variables, and when $d$ is large, the number of representations increases dramatically. 
Hence, we need a systematic procedure to choose which pair combinations should be used to describe the dependence. 
For this purpose, we introduce the concept of a vine.
Since it was invented in the field of graphical modeling \citep{bedford2002}, it is convenient to use diagrams referred to as graphs.

\subsubsection{Vine}
\begin{figure}
\centering\includegraphics[width=0.3\textwidth]{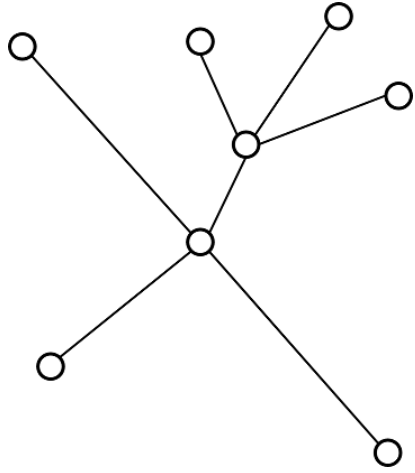}
\caption{An example of a tree graph. 
}
\label{fig:tree}
\end{figure}

In order to define it, we have to introduce some graph-theoretical terminologies. 
We start from the definition of a tree in graph theory.
\begin{definition}
{\bf (tree)}
\\
Consider a set of $d$ nodes. 
When a graph $T$ is connected and has no cycles, $T$ is a tree. 
\end{definition}
An example of a tree is presented in Fig.~\ref{fig:tree}. 

Based on the concept of tree graph, we define a vine.
\begin{definition}
{\bf (vine)}
\\
  A vine $\mathcal{V}$ on $d$ elements $\{ 1, 2, \dots, d\}$ is a set of trees $T_i\; (i = 1, \dots, d-1)$, which satisfies the following conditions: 
  \begin{enumerate}
      \item $T_1$ is a connected tree that have $\{ 1, 2, \dots, d\}$ as a set of nodes and $E_1$ as a set of edges, 
      \item For $i = 2, \dots, d-1$, $T_i$ is a tree that have $E_{i-1}$ as a set of nodes and $E_i$ as a set of edges. 
  \end{enumerate}
\end{definition}

\begin{figure}
\centering\includegraphics[width=0.6\textwidth]{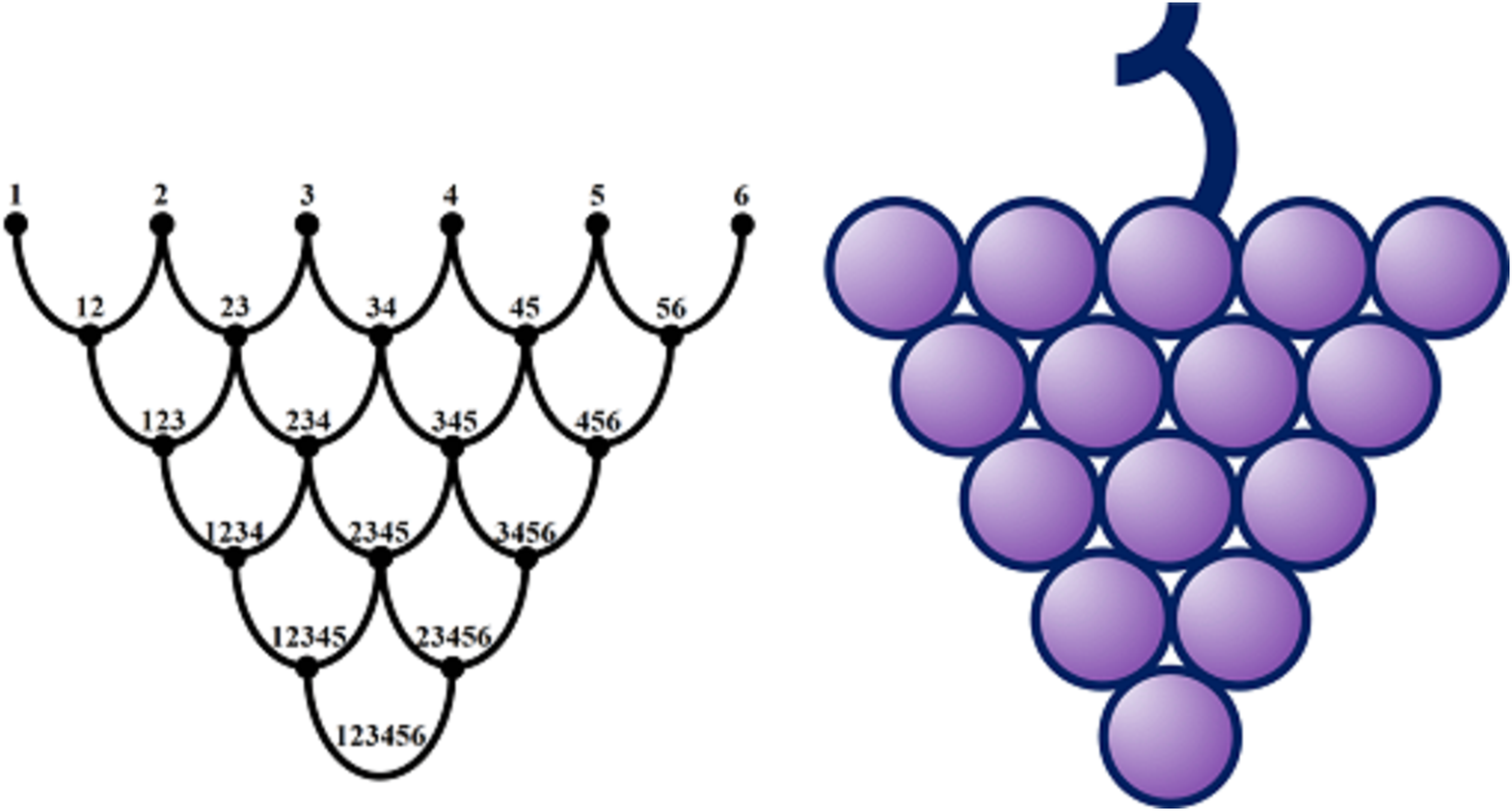}
\caption{A diagrammatic presentation of a concept of vines.
As an example, we show the case with six variables. 
This is called a vine because it looks like a grape (right panel). 
}
\label{fig:vines_structure}
\end{figure}

\begin{definition}
{\bf (regular vine)}
\\
If two nodes in tree $T_{i + 1}$ are joined by an edge, the corresponding edges in tree $i$ share a node. 
This is referred to as the proximity condition.
\end{definition}
Following discussions will be restricted to regular vines without any loss of generality, since the class of regular vines is still so large that it can treat most of the practical cases. 
The structure of a regular vine is schematically described in Fig.~\ref{fig:vines_structure}. 
The term "vine" is named after the fact that its botryoidal structure looks similar to a cluster of grapes in its appearance (see Fig.~\ref{fig:vines_structure}: e.g.  Chapter~1 of \citealt{kurowicka2011}).
We note that the tree structure is not strictly necessary for applying the pair-copula methodology, but it helps with  identifying the different pair-copula decompositions \citep{aas2009}.

\subsubsection{Frequently used vines}

In practice, two subclasses of vines are frequently used. 
They are so called "D-vine" and "C-vine", introduced as follows. 
\begin{definition}
{\bf D (drawable)-vine} \\
Any joint PDF $f(x_1, \dots, x_d)$ can be written down by D-vine as follows. 
\begin{eqnarray}
  f(x_1, \cdots, x_d ) = \prod_{j=1}^{d-1}\prod_{i=1}^{d-j} c_{i, (i+j)|(i+1), \dots, (i+j-1)}\left[F(x_i|x_i, \dots, x_{i+j-1}), F(x_{i+j}|x_{i+1}, \dots, x_{x_{i+j-1}} )\right]
  \prod^{d}_{k=1} f_k (x_k) \; ,
\end{eqnarray}
where index $j$ identifies the trees, and $i$ runs over each tree
\citep{bedford2001}. 
\end{definition}

\begin{figure}
\centering\includegraphics[width=0.45\textwidth]{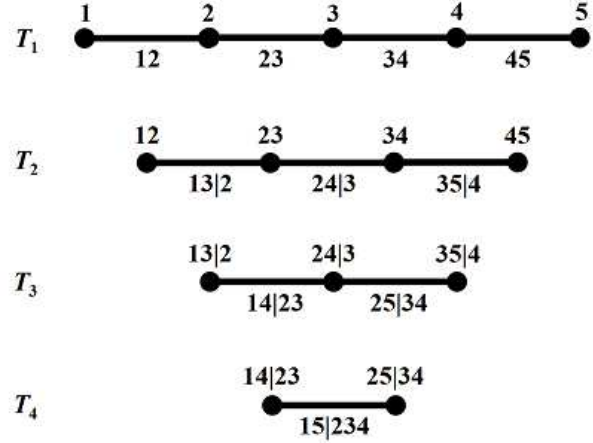}
\caption{A diagrammatic representation of a D-vine with five variables. 
It contains four trees and ten edges. 
Each edge is associated with a pair copula. 
}
\label{fig:d_vine}
\end{figure}

\noindent
The definition of D-vine means that for any tree $T_i$, the number of edges connected to each nodes never exceeds 2. 
To have a concrete idea, we present examples for $d = 3, 4, $ and $5$. 
\begin{eqnarray}
  f(x_1, x_2, x_3 ) &=& 
  c_{13|2} \left[ F(x_1|x_2)F(x_3|x_2) \right] \nonumber \\
  &&c_{12}\left[F_1(x_1), F_2(x_2) \right] c_{23}\left[F_2(x_2), F_3(x_3) \right] \nonumber \\
  &&f_1(x_1) f_2(x_2) f_3(x_3) \; , \label{eq:d_vine_3}\\
  f(x_1, x_2, x_3, x_4 ) &=& 
  c_{14|23} \left[ F(x_1|x_2,x_3)F(x_4|x_2,x_3) \right] \nonumber \\ &&c_{13|2}\left[F(x_1|x_2), F_3(x_3|x_2) \right] c_{24|3}\left[F(x_2|x_3), F(x_4|x_3) \right] \nonumber \\ 
  &&c_{12}\left[F_1(x_1), F_2(x_2) \right] 
  c_{23}\left[F_2(x_2), F_3(x_3) \right] 
  c_{34}\left[F_3(x_3), F_4(x_4) \right] \nonumber \\
  &&f_1(x_1) f_2(x_2) f_3(x_3) f_4(x_4)\; , \label{eq:d_vine_4}\\
  f(x_1, x_2, x_3, x_4, x_5) &=& 
  c_{15|234} \left[ F(x_1|x_2, x_3, x_4) F(x_5|x_2, x_3, x_4) \right] \nonumber \\
  &&c_{14|23} \left[ F(x_1|x_2, x_3)F(x_4|x_2,x_3) \right] 
  c_{25|34} \left[ F(x_2|x_3, x_4)F(x_5|x_3,x_4)\right] \nonumber \\
  &&c_{13|2}\left[F(x_1|x_2), F_3(x_3|x_2) \right] c_{24|3}\left[F(x_2|x_3), F(x_4|x_3) \right] 
  c_{35|4}\left[F(x_3|x_4), F(x_5|x_4) \right] \nonumber \\ 
  &&c_{12}\left[F_1(x_1), F_2(x_2) \right] 
  c_{23}\left[F_2(x_2), F_3(x_3) \right] 
  c_{34}\left[F_3(x_3), F_4(x_4) \right] 
  c_{45}\left[F_4(x_4), F_5(x_5) \right]\nonumber \\
  &&f_1(x_1) f_2(x_2) f_3(x_3) f_4(x_4) f_5(x_5) \; .\label{eq:d_vine_5}
\end{eqnarray}
A diagrammatic representation of a D-vine with five variables is shown in Fig.~\ref{fig:d_vine}. 
This describes the dependence structure of the D-vine well. 
In this case it has four layers of tree structure, labelled as $T_i ( i=1,\dots , 4)$. 
Each edge is associated with a pair copula. 

\begin{definition}
{\bf C (canonical)-vine} \\
Any joint PDF $f(x_1, \dots, x_d)$ can be written down by C-vine as follows. 
\begin{eqnarray}
  f(x_1, \cdots, x_d ) = \prod_{j=1}^{d-1}\prod_{i=1}^{d-j} c_{j,(j+i)|1, \dots, (j-1)}\left[F(x_j|x_1, \dots, x_{j-1}), F(j_i|x_1, \dots, x_{j-1})\right]
  \prod^{d}_{k=1} f_k (x_k) \; .
\end{eqnarray}
Each tree $T_j$ has a unique node connected to $d-j$ edges. 
\end{definition}

\noindent
When we know a particular variable is a key that governs the interaction in the dataset, the C-vine has a great advantage. 
We can decide this "pivot" variable at the root of the C-vine. 
We show examples for $d = 3, 4, $ and $5$. 
\begin{eqnarray}
  f(x_1, x_2, x_3 ) &=& c_{23|1} \left[ F(x_2|x_3)F(x_3|x_1) \right] \nonumber \\
  &&c_{12}\left[F_1(x_1), F_2(x_2) \right] 
  c_{13}\left[F_1(x_1), F_3(x_3) \right] \nonumber \\
  &&f_1(x_1) f_2(x_2) f_3(x_3) \; ,\label{eq:c_vine_3} \\
  f(x_1, x_2, x_3, x_4 ) &=& 
  c_{34|12} \left[ F(x_3|x_1, x_2)F(x_4|x_1,x_2) \right] \nonumber \\
  &&c_{23|1}\left[F(x_2|x_1), F(x_3|x_1) \right] c_{24|1}\left[F(x_2|x_1), F(x_4|x_1) \right] \nonumber \\ 
  &&c_{12}\left[F_1(x_1), F_2(x_2) \right] 
  c_{13}\left[F_1(x_1), F_3(x_3) \right] 
  c_{14}\left[F_1(x_1), F_4(x_4) \right] \nonumber \\
  &&f_1(x_1) f_2(x_2) f_3(x_3) f_4(x_4)\; , \label{eq:c_vine_4}\\
  f(x_1, x_2, x_3, x_4, x_5) &=& 
  c_{45|123} \left[ F(x_4|x_1, x_2, x_3) F(x_5|x_1, x_2, x_3) \right] \nonumber \\
  &&c_{34|12} \left[ F(x_3|x_1, x_2)F(x_4|x_1,x_2) \right] 
  c_{35|12} \left[ F(x_3|x_1, x_2)F(x_5|x_1,x_2)\right] \nonumber \\
  &&c_{23|1}\left[F(x_2|x_1), F(x_3|x_1) \right] c_{24|1}\left[F(x_2|x_1), F(x_4|x_1) \right] 
  c_{25|1}\left[F(x_2|x_1), F(x_5|x_1) \right] \nonumber \\ 
  &&c_{12}\left[F_1(x_1), F_2(x_2) \right] 
  c_{13}\left[F_1(x_1), F_3(x_3) \right] 
  c_{14}\left[F_1(x_1), F_4(x_4) \right] 
  c_{15}\left[F_1(x_1), F_5(x_5) \right] \nonumber \\
  &&f_1(x_1) f_2(x_2) f_3(x_3) f_4(x_4) f_5(x_5)\; .\label{eq:c_vine_5}
\end{eqnarray}

\begin{figure}
\centering\includegraphics[width=0.45\textwidth]{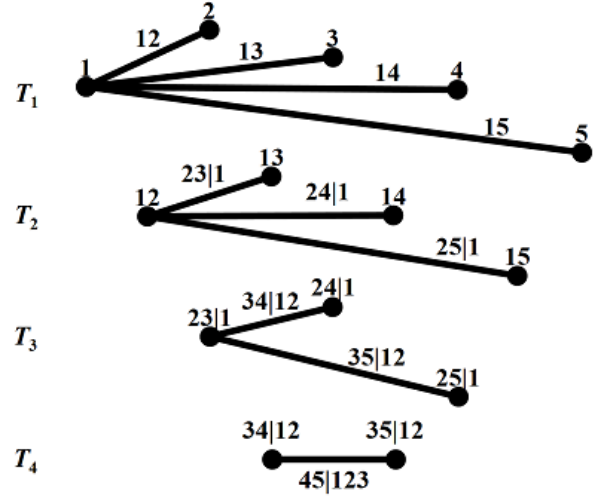}
\caption{Same as Fig.~\ref{fig:d_vine} but for a C-vine with five variables. 
}
\label{fig:c_vine}
\end{figure}

For the case of $d=3$, the general expression for vine structure is expressed as eq.~(\ref{eq:d_vine_3}) or (\ref{eq:c_vine_3}). 
It is valid for both D- and C-vines. 
There are six ways with a permutation between $x_1, x_2$, and $x_3$, but only three of them yield different pair-copula decompositions. 
Further, each of the three correspond both to D- and C-vine. 
Namely, both D- and C-vines cover the whole possible structures of pair copula decompositions for $d=3$. 
For $d=4$, there are 24 regular vine decompositions, of which 12 are D-vine and 12 are C-vine ones. 
There is no overlap between any of the D- and C-vines. 
Further, there are no other regular vine decompositions. 
This guarantees the wide range of applicability of D- and C-vines for the pair-copula decomposition. 
For $d = 5$, there are 60 different D-vines and 60 different C-vines. 
Between any of these 60 D-vines and 60 C-vines, there is no overlap. 
However, unlike $d \leq 4$, there are 120 more regular vines that are not D- nor C-vines. 
Hence in total there are 240 different possible pair-copula decompositions. 
In general, for any $d$, there are $d!/2$ D-vines and the same number of C-vines.
Detailed explanations on these results are presented in \citet{aas2009}. 

\subsection{Likelihood for vines}

Practically, we can safely restrict the pair-copula decompositions to D- and C-vines. 
When we have $n$ data sample $\{\vec{x}_{m}\} = \{(x_{m,1}, x_{m,2}, \dots, x_{m,d} )\}$ $(m = 1, \dots, n)$, the log likelihood of a D-vine for the parameter estimation is
\begin{eqnarray}\label{eq:likelihood_d_vine}
  \ln \mathcal{L}(\vec{\theta}; \vec{x}_{m}, m = 1, \dots, n) %\nonumber \\
  &=&
  \sum_{m=1}^{n} \ln f(\vec{\theta}; \vec{x}_{m}) \nonumber \\
  &=& \sum_{m=1}^{n} \sum_{j=1}^{d-1}\sum_{i=1}^{d-j} \ln %c(\vec{\theta}_{i, (i+j)|(i+1), \dots, (i+j-1)})_{i, (i+j)|(i+1), \dots, (i+j-1)}
  c[\vec{\theta}(\mbox{copula})]_{i, (i+j)|(i+1), \dots, (i+j-1)}
  \left[F(x_i|x_i, \dots, x_{i+j-1}), F(x_{i+j}|x_{i+1}, \dots, x_{x_{i+j-1}} )\right] \nonumber \\
  &&\quad 
  +\sum^{d}_{k=1} \ln f_k [\vec{\theta}(\mbox{marginal}); x_k] \; ,
%  &&\ln \mathcal{L}(\vec{\theta}; \vec{x}_{m}, m = 1, \dots, n) \nonumber \\
%  &&=
%  \sum_{m=1}^{n} \ln f(\vec{\theta}; \vec{x}_{m}) \nonumber \\
%  &&= \sum_{m=1}^{n} \sum_{j=1}^{d-1}\sum_{i=1}^{d-j} \ln %c(\vec{\theta}_{i, (i+j)|(i+1), \dots, (i+j-1)})_{i, (i+j)|(i+1), \dots, (i+j-1)}
%  c[\vec{\theta}(\mbox{copula})]_{i, (i+j)|(i+1), \dots, (i+j-1)}
%  \left[F(x_i|x_i, \dots, x_{i+j-1}), F(x_{i+j}|x_{i+1}, \dots, x_{x_{i+j-1}} )\right] \nonumber \\
%  &&\quad 
%  +\sum^{d}_{k=1} \ln f_k [\vec{\theta}(\mbox{marginal}); x_k] \; ,
\end{eqnarray}
where we denote the parameter vectors for copulae and marginals in a symbolic way for saving space. 
Similarly, the log likelihood of a C-vine is
\begin{eqnarray}\label{eq:likelihood_c_vine}
  \ln \mathcal{L}(\vec{\theta}; \vec{x}_{m}, m = 1, \dots, n) %\nonumber \\
%  &=& \sum_{m=1}^{n} \ln f(\vec{\theta}; \vec{x}_{m}) \nonumber \\
  &=& \sum_{m=1}^{n} \sum_{j=1}^{d-1}\sum_{i=1}^{d-j} \ln 
  c[\vec{\theta}(\mbox{copula})]_{j,(j+i)|1, \dots, (j-1)}
  \left[F(x_i|x_i, \dots, x_{i+j-1}), F(x_{i+j}|x_{i+1}, \dots, x_{x_{i+j-1}} )\right] \nonumber \\
  &&\quad 
  +\sum^{d}_{k=1} \ln f_k [\vec{\theta}(\mbox{marginal}); x_k] \; .
\end{eqnarray}
We should maximize the log likelihood eq.~(\ref{eq:likelihood_d_vine}) or (\ref{eq:likelihood_c_vine}) to estimate the parameter set $\vec{\theta} = [\vec{\theta}(\mbox{copula}), \vec{\theta}(\mbox{marginal})]$. 
In principle, both $\vec{\theta}(\mbox{copula})$ and $ \vec{\theta}(\mbox{marginal})$ can be estimated simultaneously. 
In the research fields such as economics, however, the likelihood estimation of the marginals is often found to be difficult or even implausible.
Hence, they do not use the exact form of the likelihood but instead maximize the so-called pseudo-likelihood \citep{aas2009}. 
In contrast, in astrophysics, we have a rich field of research on the estimation of the marginals, e.g. the LF or MF of galaxies at a certain observed wavelength \citep[e.g.][]{takeuchi00b,johnston2011}.
Then, we can simply use the estimated marginals and plug in them for the log-likelihood. 
Namely, we can omit the estimation step for the marginals when we try to estimate copula parameters in a different sense from other research fields. 

On this step, we can obtain the error of each parameter and the goodness of fit indicator(s) of the model, since we have the likelihood ellipsoid and maximum likelihood value.

\subsection{Model selection with vines}

\begin{figure}
\centering\includegraphics[width=0.6\textwidth]{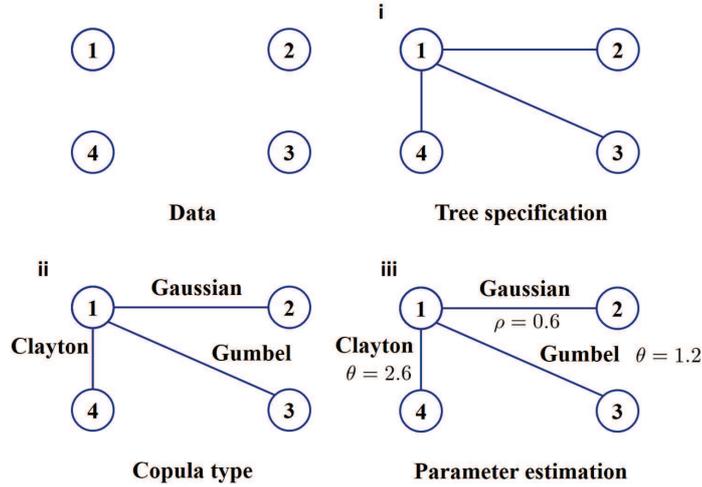}
\caption{A schematic example for the description of the estimation with vines. 
Gaussian, Clayton, and Gumbel stand for three popular copula types often used in practice. 
Each of them has some specifying parameters, and they are estimated by usual statistical procedure in Step iii. 
}
\label{fig:estimation}
\end{figure}

The likelihood estimation we discussed above is only one of the steps of the full estimation problem. 
Schematically, the model to be specified has a following structure as
\begin{eqnarray}
  \mbox{Model} = \mbox{structure (trees)} + \mbox{copula families} + \mbox{copula parameters} \; .
\end{eqnarray}
Namely, for the estimation procedure we should consider
\begin{enumerate}
    \item selection of a specific decomposition, 
    \item choice of pair-copula types, 
    \item estimation of the copula parameters.
\end{enumerate}
This is schematically described in Fig.~\ref{fig:estimation}. 
As for the copula types, since we have an infinitely large degree of freedom for the choice of copulae, the model selection is fundamentally important. 
Often we have to choose an optimal model from large choice of candidate models with different number of parameters. 
In such a case, usual goodness-of-fit method does not work, since obviously more parameters give a better fit. 
For such a case, a model selection procedure should be used instead .
The most popular tool for the model selection is the information criterion \citep[e.g.][and references therein]{takeuchi00a}. 
One of the first and most widely used information criterion is the Akaike Information Criterion (AIC: \citealt{akaike1974}), 
\begin{eqnarray}
  \mbox{AIC}(q) = -2[\ln \mathcal{L}(\hat{\vec{\theta}}) - q]
\end{eqnarray}
where $\hat{\vec{\theta}}$ stands for the parameter that maximizes the likelihood, and $q$ is the number of parameters. 
The model that gives the smallest AIC is selected. 
This is a natural extension of the classical maximum likelihood estimation, corrected for the bias introduced by the parameter estimation step \citep{akaike1974}. 
Some other information criteria are also used.
Among them, the Bayesian Information Criterion (BIC) 
\begin{eqnarray}
  \mbox{BIC}(q) = -2\left[\ln \mathcal{L}(\hat{\vec{\theta}}) - \frac{q}{2}\ln n\right]
\end{eqnarray}
($n$: sample size) is also often used \citep{schwarz1978}.
This model selection is performed in the step of copula selection. 
Each copula is determined by the evaluation of such information criterion among all possible copula types, 
as well as the classical goodness-of-fit indicators like $\chi^2$-statistic. 

As we saw in Section~\ref{subsec:vine_copula}, the tree determination and subsequent specification of copulae would be computationally heavy because of the factor $d!/2$. 
Thanks to the present-day development of software, we can treat a problem with a dimension up to $d \sim 500$.
Some software packages are available for this problem \citep[e.g.][]{brechmann2013}. 
The construction and estimation of a multivariate PDF based on the vine copula is completed by this step. 

\section{Application to construct a multivariate mass function (MF) of galaxies}
\label{sec:mlf}

The relation between stars, atomic gas, and molecular gas mass is one of the most important issues in galaxy evolution, and studied very extensively. 
For this aim, a multivariate MF is obviously a fundamental tool 
\citep[see e.g. an elaborate work of][]{rodriguez_puebla2020}.
Here we present one astrophysically interesting example of multivariate mass function, a three-variate mass function of stellar, atomic gas, and molecular gas mass. 
The formal procedure is exactly the same for the case of luminosity functions. 

\subsection{Multivariate MF}

First we prepare some notations for the multivariate MF. 
A mass function of galaxies is defined as a number density of galaxies whose mass lies between a logarithmic interval\footnote{We denote $\log x \equiv \log_{10} x$ and $\ln x \equiv \log_e x$.}
$[\log M, \log M + \pd\log M]$: 
\begin{eqnarray}
 \phi^{(1)} (M) \equiv \frac{\pd n}{\pd \log M} \;.
\end{eqnarray}
For mathematical simplicity, we define the  as {\sl being normalized}, i.e., 
\begin{eqnarray}\label{eq:normalization}
 \int \phi^{(1)} (M) \pd\log M = 1\;.
\end{eqnarray}
Hence, this corresponds to a PDF.
We also define the cumulative MF as
\begin{eqnarray}
 \Phi^{(1)} (M) \equiv \int_{\log M_{\rm min}}^{\log M} \phi^{(1)} (M') \pd \log M' \;,
\end{eqnarray}
where $M_{\rm min}$ is the minimum mass of galaxies considered.
This corresponds to the DF.

If we denote univariate MFs as $\phi^{(1)}_k(M_k)$ $(k=1, \dots, d)$,  
the joint multivariate PDF $\phi^{(d)}(M_1, \dots, L_d)$ is described by a differential copula $c(u_1, \dots, u_k)$ as
\begin{eqnarray}
  \phi^{(d)} (M_1, \dots, M_d) \hspace{-3mm}&\equiv & \hspace{-3mm}
    c \left[\Phi^{(1)}_1(M_1), \dots, \Phi^{(1)}_{d}(M_d)\right] \phi^{(1)}_1(M_1)\cdots \phi^{(1)}_d(M_d)\;.
\end{eqnarray}

For this analysis, we made use of the R package {\tt VineCopula}\footnote{https://github.com/tnagler/VineCopula.} \citep{aas2009}. 
It provides statistical inference of C- and D-vine copulae. 
This package enables us to construct copula density and tree structures from multivariate data. 
The optimal combination of copulae and their parameters are chosen through AIC, BIC and maximum likelihood estimation with {\tt RVineStructureSelect} function. 

\subsection{The stellar--atomic gas--molecular gas multivariate MF}
\label{subsec:Mstar_MHI_MH2_MF}

As marginals of the multivariate MF, we should determine the univariate MFs for stellar mass, atomic gas mass, and molecular gas mass, respectively. 
The univariate MF for atomic gas mass, and molecular gas mass are known to be well described by the Schechter function \citep{schechter76}.
\begin{eqnarray}\label{eq:schechter}
  \phi^{(1)} (M) = (\ln 10)\; \phi_* \left( \frac{M}{M_*} \right)^{1-\alpha}
  \exp \left[-\left(\frac{M}{M_*}\right)\right]\;,
\end{eqnarray}
For the atomic gas mass, we took the parameters from \citet{jones2018} but in a normalized form with eq.~(\ref{eq:normalization}), i.e., we did not use the normalization factor $\phi_{\rm HI *}$.
Similarly, we took the Schechter function parameters for the molecular gas from \citet{keres2003}. 
These parameters are summarized in Table~\ref{tab:gasMF}.

\begin{table} 
\begin{center} 
\caption{Parameters for the gas mass function} \label{tab:gasMF}
\begin{tabular}{lcccl} 
\hline \hline 
MF  & $\alpha$ & $\phi_*$ & $M_*$ & Reference\\
~ & ~ & $[\mbox{Mpc}^{-3}\mbox{dex}^{-1}]$ & $[M_\odot]$ & ~\\ 
\hline 
Atomic gas & $-1.25 \pm 0.02$ & $(4.5 \pm 0.2) \times 10^{-3}$ & $8.7 \times 10^9$ & \citet{jones2018} \\
Molecular gas & $-1.30 \pm 0.16$ & $(5.9 \pm 2.8) \times 10^{-3}$ & $9.4 \times 10^9$ & \citet{keres2003} \\
\hline\\
\end{tabular}\\ 
\end{center} 
\end{table} 

Recent studies revealed that the stellar MF is, however, better described by a double Schechter function \citep[e.g.][]{dsouza2015}
\begin{eqnarray}\label{eq:double_schechter}
  \phi^{(1)} (M) = 
  \frac{\phi_{1*}}{M_{1*}} \left( \frac{M}{M_{1*}} \right)^{-\alpha_1}
  \exp \left[-\left(\frac{M}{M_{1*}}\right)\right] + 
  \frac{\phi_{2*}}{M_{2*}} \left( \frac{M}{M_{2*}} \right)^{-\alpha_2}
  \exp \left[-\left(\frac{M}{M_{2*}}\right)\right] 
  \;.
\end{eqnarray}
We should note that \citet{dsouza2015} defined the double Schechter function (eq.~(\ref{eq:double_schechter})) for a linear mass interval $\pd M$, {\sl not} $\pd \log M$. 
The parameters are shown in Table~\ref{tab:stellarMF}. 

\begin{table} 
\begin{center} 
\caption{Parameters for the stellar mass function} \label{tab:stellarMF}
\begin{tabular}{ccccccl} 
\hline \hline 
  $\alpha_1$ & $\phi_{_1*}$ & $M_{1*}$ & $\alpha_2$ & $\phi_{_2*}$ & $M_{2*}$ & Reference \\
  ~ & $[\mbox{Mpc}^{-3}\mbox{dex}^{-1}]$ & $[M_\odot]$ & ~ & $[\mbox{Mpc}^{-3}\mbox{dex}^{-1}]$ & $[M_\odot]$ & \\ 
\hline
1.082 & $6.0 \times 10^{-2}$ & $4.1 \times 10^{10}$ & $1.120$ & $2.5 \times 10^{-3}$ & $9.9 \times 10^{10}$ & \citet{dsouza2015} \\
\hline\\
\end{tabular}\\ 
\end{center} 
\end{table}

\subsection{Data}

In this work, we used a subsample of the combined dataset compiled by \citet{calette2018}. 
Their original sample consists of Golden, Silver, and Bronze Categories both for H{\sc i} and H$_2$. 
Full details of the original sample are are found in Appendix of \citet{calette2018}. 
We briefly describe the dataset used here. 

\subsubsection{The compiled galaxy sample with H{\sc i} information}

%The galaxy dataset with H{\sc i} information are classified into three categories as follows: 
We used the following datasets for H{\sc i} information. 

\noindent
{\bf Golden Category}
\begin{itemize}
    % \item Updated Nearby Galaxy Catalog (UNGC; \citealt{karachentsev2013, karachentsev2014}): 
    % a sample of 869 galaxies in the Local Volume, located within 11~Mpc. 
    % The sample is originally complete to $M_B < -11$~mag, spanning all morphologies, but \citet{calette2018} restricted it with a more conservative limit considering the surface brightness limit. 
    \item {\sl {\sl GALEX}} Arecibo SDSS Survey (GASS; \citealt{catinella2013}): 
    an optically-selected subsample of 760 galaxies more massive than $10^{10}\;M_\odot$ taken from a parent SDSS DR6 sample volume limited in the redshift range $0.025 < z < 0.05$ and cross-matched with the ALFALFA and {\sl {\sl GALEX}} surveys.
    \item Field galaxies from the Herschel Reference Survey (HRS; \citealt{boselli2010,boselli2014a,boselli2014b,boselli2014c}): 
    a $K$-band volume limited ($15 \leq D\; [\mbox{Mpc}] \leq 25$) sample of 323 galaxies complete to $K_{\rm s} = -12$ and $-8.7$~mag for late type galaxies and early type galaxies, respectively. 
    \item Field early type galaxies from the ATLAS$^{\rm 3D}$ HI sample \citep{serra2012}: 
    a sample of 166 local early type galaxies observed in detail with integral field unities (IFUs; Cappellari et al. 2011). 
    The distance range of the sample is between 10 and 47~Mpc; the sample includes 39 galaxies from the Virgo Cluster, but for the Golden category, the early type galaxies  in the Virgo cluster core were excluded by \citet{calette2018}. 
\end{itemize}

\begin{comment}
\noindent
{\bf Silver Category}
\begin{itemize}
     \item Nearby Field Galaxy Survey (NFGS; \citealt{jansen2000a,jansen2000b, wei2010, kannappan2013}, and references therein): 
    a broadly representative sample of 198 local galaxies spanning stellar masses $M_* \simeq 10^8 \mbox{--} 10^{12}\; M_\odot$ and all morphological types.
    \item \citet{stark2013} compilation: 
    compiled from the literature and homogenized 323 galaxies with available HI, CO, and multi-band imaging data. Most of the compiled galaxies are from the GASS, NFGS and ATLAS$^{\rm 3D}$ surveys described above. 
    \citet{calette2018} used 67 galaxies not in these surveys.
    \item \citet{leroy2008} THINGS sample: 
    a sample of 23 nearby, star-forming galaxies, which we associate with late type galaxies. 
    \item Dwarf late type galaxies \citep{geha2006}: 
    a sample of 101 dwarf galaxies, 88 out of them with H{\sc i} measurements and being of late type.
    \item ALFALFA dwarf sample \citep{huang2012}: 
    It consists of 176 low HI mass dwarf galaxies from the ALFALFA survey. The galaxies were selected to have $M_{\rm HI} < 10^{7.7} \; M_\odot$ and H{\sc i} line widths $< 80 \mbox{km}\, \mbox{s}^{-1}$.
\end{itemize}
\end{comment}

\noindent
{\bf Bronze Category}
\begin{itemize}
    % \item UNAM-KIAS catalog of isolated galaxies \citep{hernandez_toledo2010}: 
    % a magnitude limited sample ($m_r < 15.2$~mag) of galaxies from the SDSS DR5 that fulfill strict isolation criteria; it is composed of 1520 galaxies spanning all morphological types.
    \item Analysis of the interstellar Medium of Isolated GAlaxies (AMIGA; \citealt{lisenfeld2011}): 
    a redshift-limited sample ($1500 \leq v_{\rm rec} [\mbox{km}\, \mbox{s}^{-1}] \leq 5000$) consisting of 273 isolated galaxies with reported multi-band imaging and CO data. 
    % \item Low-mass Isolated galaxies \citep{bradford2015}: 
    % a sample of 148 isolated low-mass galaxies ($7 \leq \log M_*/[M_\odot]) \leq 9.5$) drawn from the SDSS NSA catalog. 
    % \item Herschel Reference Survey -- Virgo galaxies: 
    % the same HRS sample described above but including only galaxies from the Virgo Cluster central regions A and B (59 galaxies).
    % \item ATLAS$^{3D}$ H{\sc i} sample -- Virgo core early type galaxies: 
    % the same ATLAS3D sample described above but taking into account account only the Virgo core early type galaxies (15 galaxies).
\end{itemize}

\subsubsection{The compiled galaxy sample with CO (H$_2$) information}

%The galaxy dataset with CO (H$_2$) information are classified into three categories as follows: 
We used the following datasets for H{\sc i} information. 

\noindent
{\bf Golden Category}
\begin{itemize}
    \item Field galaxies from the Herschel Reference Survey (HRS): 
    the same sample described above (excluding Virgo Cluster core), with 155 galaxies with available CO information (101 detections and 54 non-detections). 
    \item CO Legacy Legacy Database for GASS (COLD GASS; \citealt{saintonge2011}): 
    a program aimed at observing CO(1--0) line fluxes with the IRAM 30 m telescope for galaxies from the GASS survey described above. From the CO fluxes, the total CO luminosities, (and hence the H$_2$  masses) were calculated for 349 galaxies.
    \item Field early type galaxies from the ATLAS$^{\rm 3D}$ H$_2$ sample \citep{young2011}: 
    the same sample described above (excluding the Virgo Cluster core) but with observations in CO using the IRAM 30 m Radio Telescope. The sample amounts for 243 early type galaxies with CO observations. 
\end{itemize}

\begin{comment}
\noindent
{\bf Silver Category}
\begin{itemize}
    \item \citet{stark2013} compilation: 
     the same compiled galaxy sample described above. 
     The authors observed 35 galaxies of the NFGS with the IRAM 30 m and the ARO 12 m telescopes to measure the CO ($J$: 2--1) (IRAM) and ($J$: 1--0) (IRAM and ARO) lines. 
     For the other galaxies, the H2 information from previous works was used. 
     \item \citet{leroy2008} HERACLES sample: 
     the same sample described above. 
     The H$_2$ information for the 23 late type galaxies comes from the CO J: 2--1 maps from the HERA CO-Line Extragalactic Survey HERACLES \citep{leroy2008}. 
    \item APEX Low-redshift Legacy Survey for MOlecular Gas (ALLSMOG; \citealt{bothwell2014}): 
    sing the APEX telescope, the CO(2--1) emission line was measured to trace H$_2$ in 42 late type galaxies with masses $8.5 <log(M_*/[M_\odot])< 10$, at $0.01 < z < 0.03$ and with metallicities $12 + log(\mbox{O}/\mbox{H}) > 8.5$. 
    \item \citet{bauermeister2013} compilation: 
    Eight galaxies in the low-redshift range $0.05 \leq z \leq 0.1$. 
\end{itemize}
\end{comment}

\noindent
{\bf Bronze Category}
\begin{itemize}
    \item Analysis of the interstellar Medium of Isolated GAlaxies (AMIGA; \citealt{lisenfeld2011}: 
    the same sample described above.
    The authors carried out their own observations of CO($J$: 1--0) with the IRAM 30 m or the 14 m FCRAO telescopes for 189 galaxies; 87 more were compiled from the literature. 
    % \item Herschel Reference Survey -- Virgo galaxies: 
    % the same as above (62 galaxies).
    % \item ATLAS$^{\rm 3D}$ H{\sc i} sample -- Virgo core early type galaxies: 
    % the same as above (21 galaxies).
\end{itemize}

\subsection{Result}
\label{subsec:result}

\begin{figure}
\centering\includegraphics[width=0.9\textwidth]{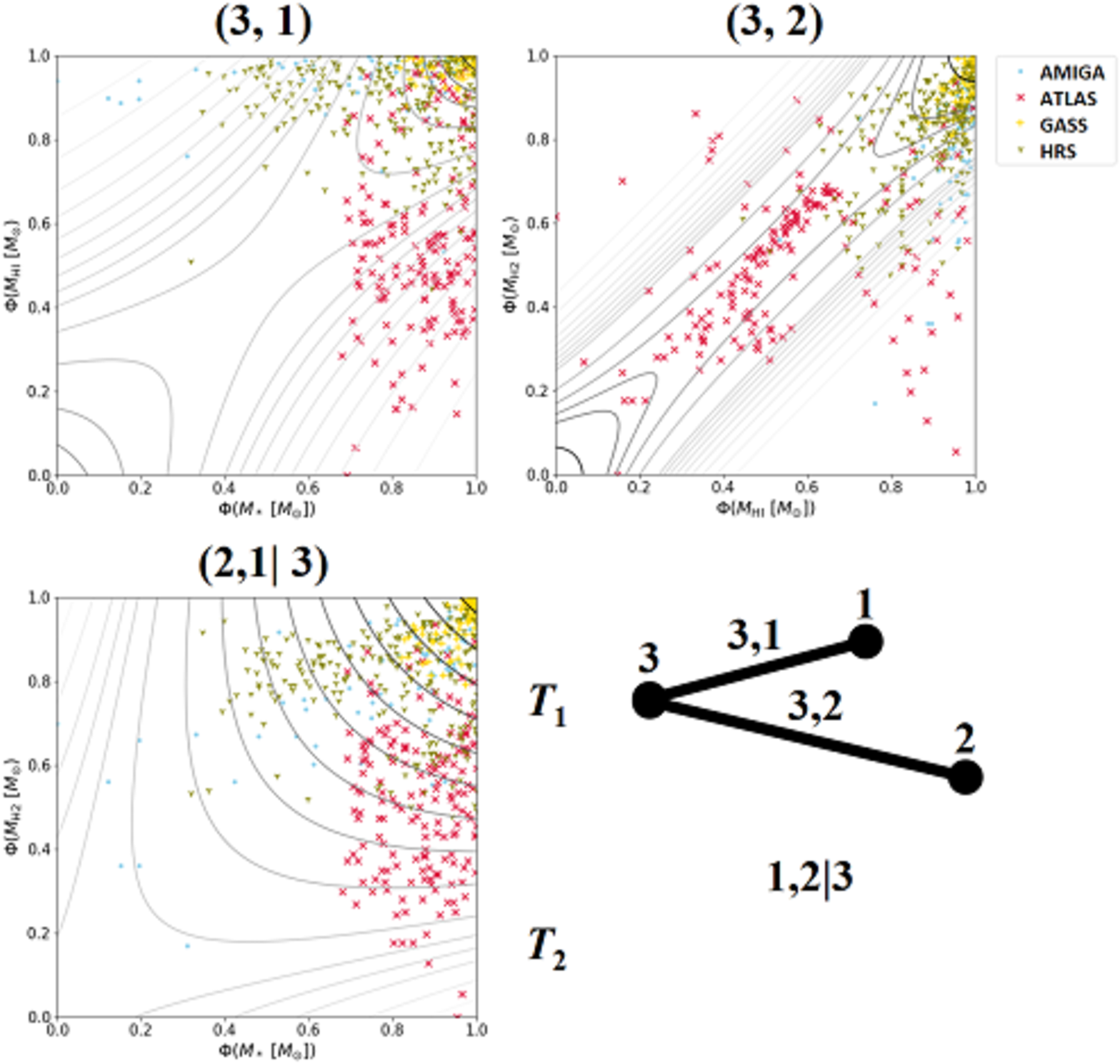}
%\centering\includegraphics[height=0.4\textwidth]{copula_Mstar_MHI.PNG}
%\centering\includegraphics[height=0.4\textwidth]{copula_MHI_MH2.PNG}
%\centering\includegraphics[height=0.4\textwidth]{copula_Mstar_MH2.PNG}
\caption{Vine copulae estimated from the $M_*$-$M_{\rm HI}$-$M_{{\rm H}_2}$ data. 
Top-left: the estimated copula for $M_{\rm H_2}$--$M_*$ relation, 
Top-right: the estimated copula for $M_{\rm H_2}$--$M_{\rm HI}$ relation, 
Bottom-left: the estimated conditional copula for $M_*$--$M_{\rm HI}$ relation with $M_{{\rm H}_2}$ given, 
and Bottom-right: tree structure of vine copulae. 
The structure of the estimated C-vine is labelled as (3,1), (3,2), and (2, 1| 3), where node 1 corresponds to $M_*$, node 2 to $M_{\rm HI}$, and node 3 to $M_{\rm H_2}$, respectively.
}
\label{fig:vine_copula_Mstar_MHI_MH2}
\end{figure}

We present the constructed $M_*$--$M_{\rm HI}$--$M_{\rm H_2}$ MF with a 3-dim vine copula.  
The goodness-of-fit analysis of the obtained 3-dim MF to the dataset gave the $p$-value $p = 0.41$ that suggests that the fit is appropriate. 
The $M_*$--$M_{\rm HI}$--$M_{\rm H_2}$ MF is described by a C-vine copula. 
The structure is presented in Fig.~\ref{fig:vine_copula_Mstar_MHI_MH2}. The structure of the estimated C-vine is labelled as (3,1), (3,2), and (2,1| 3) in Fig.~\ref{fig:vine_copula_Mstar_MHI_MH2}. 
Here, 1 corresponds to $M_*$, 2 to $M_{\rm HI}$, and 3 to $M_{\rm H_2}$, respectively.

The relation between $M_*$ and $M_{H_2}$ was well described by the BB8 copula
\begin{eqnarray}
    C(u_1,u_2;\theta,\delta)=\frac{1}{\delta}
    \left\{1-{\left[1-\frac{1}{1-(1-\delta)^{\theta}}\left (1-(1-\delta u_1)^{\theta}\right ) \left(1-(1-\delta u_2)^{\theta}\right)\right]}^{\frac{1}{\theta}}\right\}, \quad (\theta \geq 1,\ \delta \in (0,1]);,
\end{eqnarray}
with parameters $\theta=6.00\pm 1.76$ and $\delta=0.49\pm 0.10$. 
Kendall's $\tau$ of this pair is $0.35$, reflecting the broad distribution with a loose correlation.

\begin{figure}
\centering\includegraphics[width=0.45\textwidth]{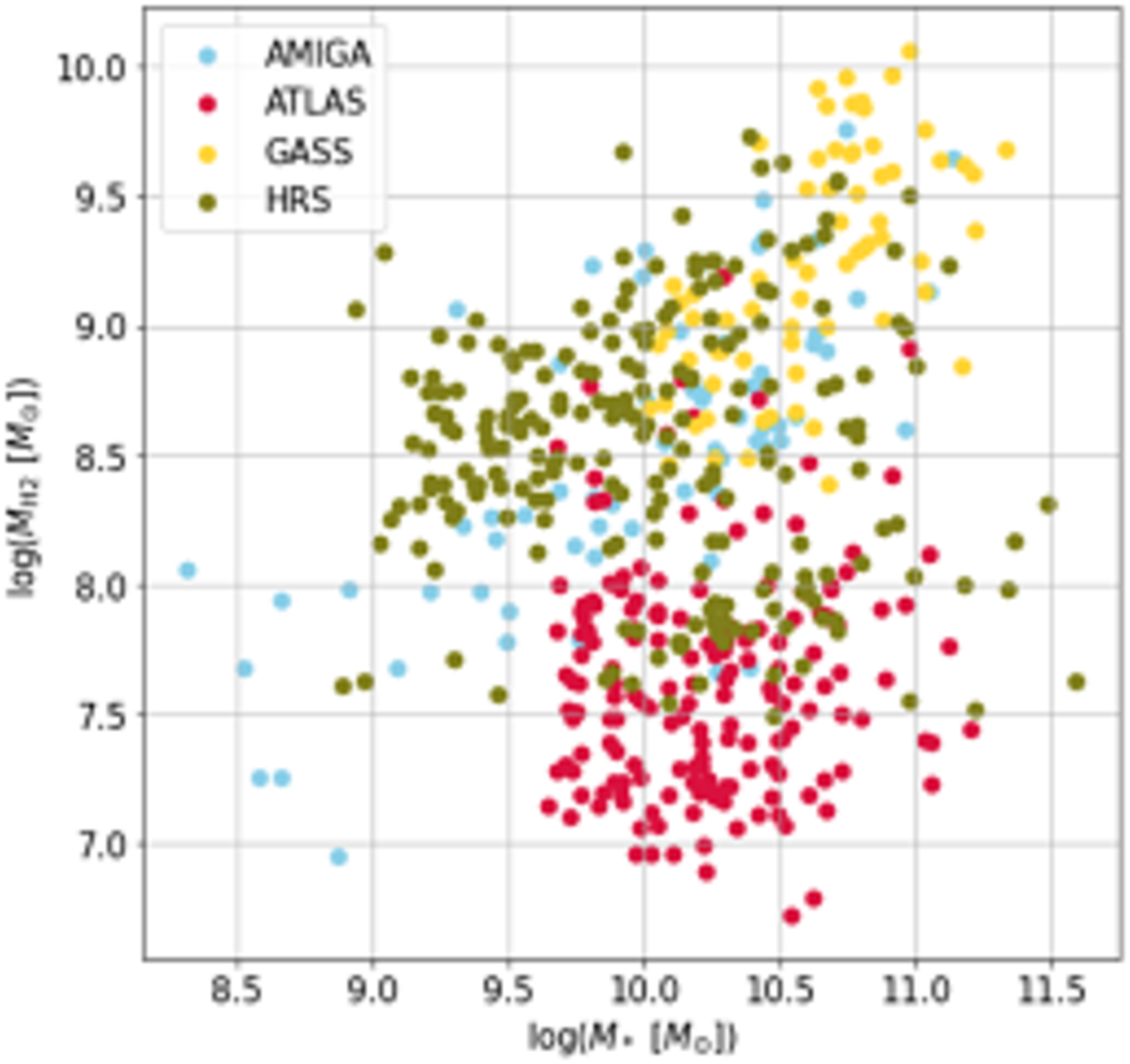}
\centering\includegraphics[width=0.48\textwidth]{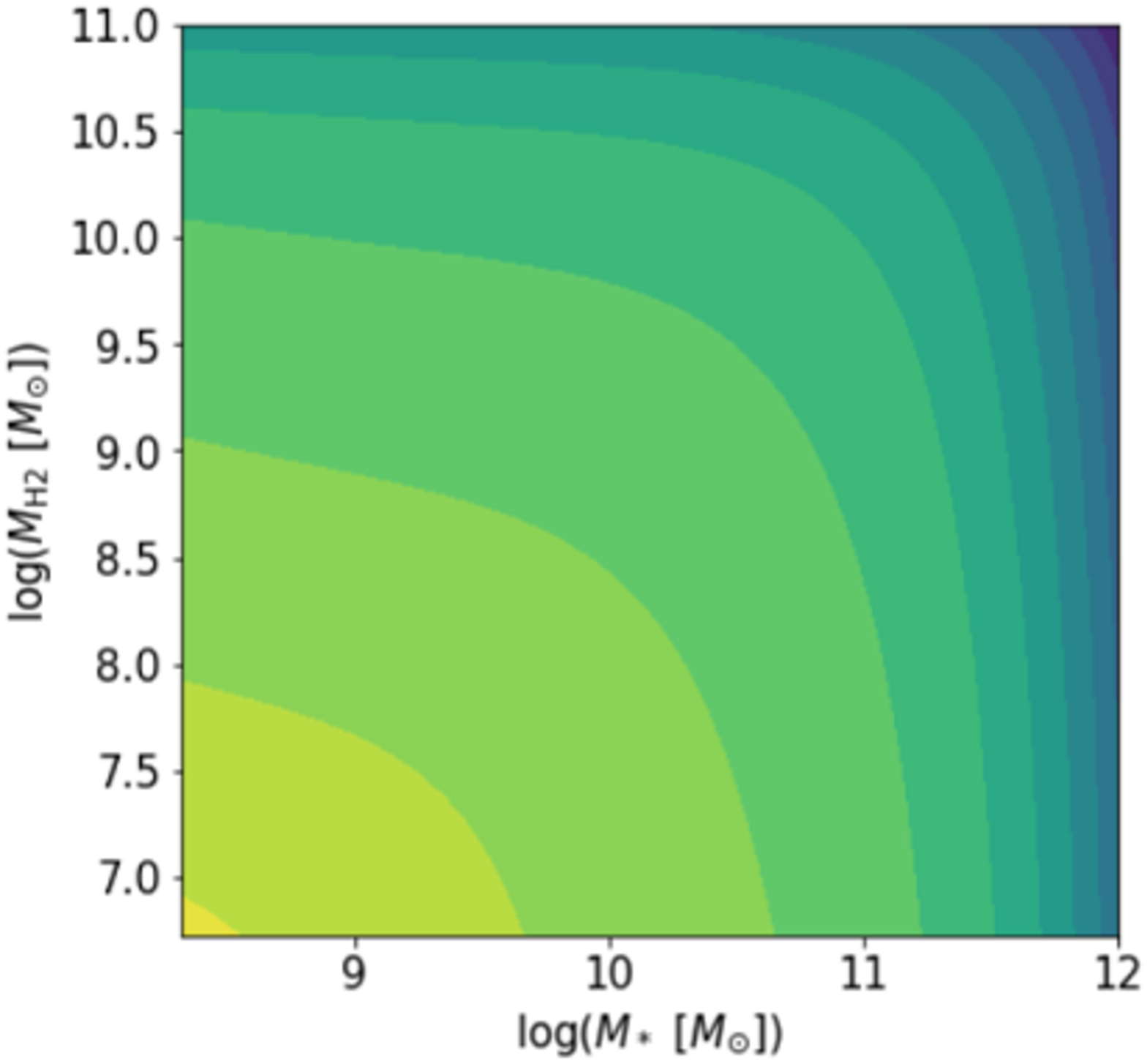}
\caption{Relation between $M_*$ and $M_{{\rm H}_2}$. 
Left: the data distribution on the  $M_*$--$M_{{\rm H}_2}$ plane. 
Right: estimated bivariate PDF. 
The selected copula is the BB8 copula with a parameter of $\theta=6.00\pm 1.76$ and $\delta=0.49\pm0.10$.
Data are taken from \citet{calette2018}. 
}
\label{fig:Mstar_MH2}
\end{figure}

\begin{figure}
\centering\includegraphics[width=0.46\textwidth]{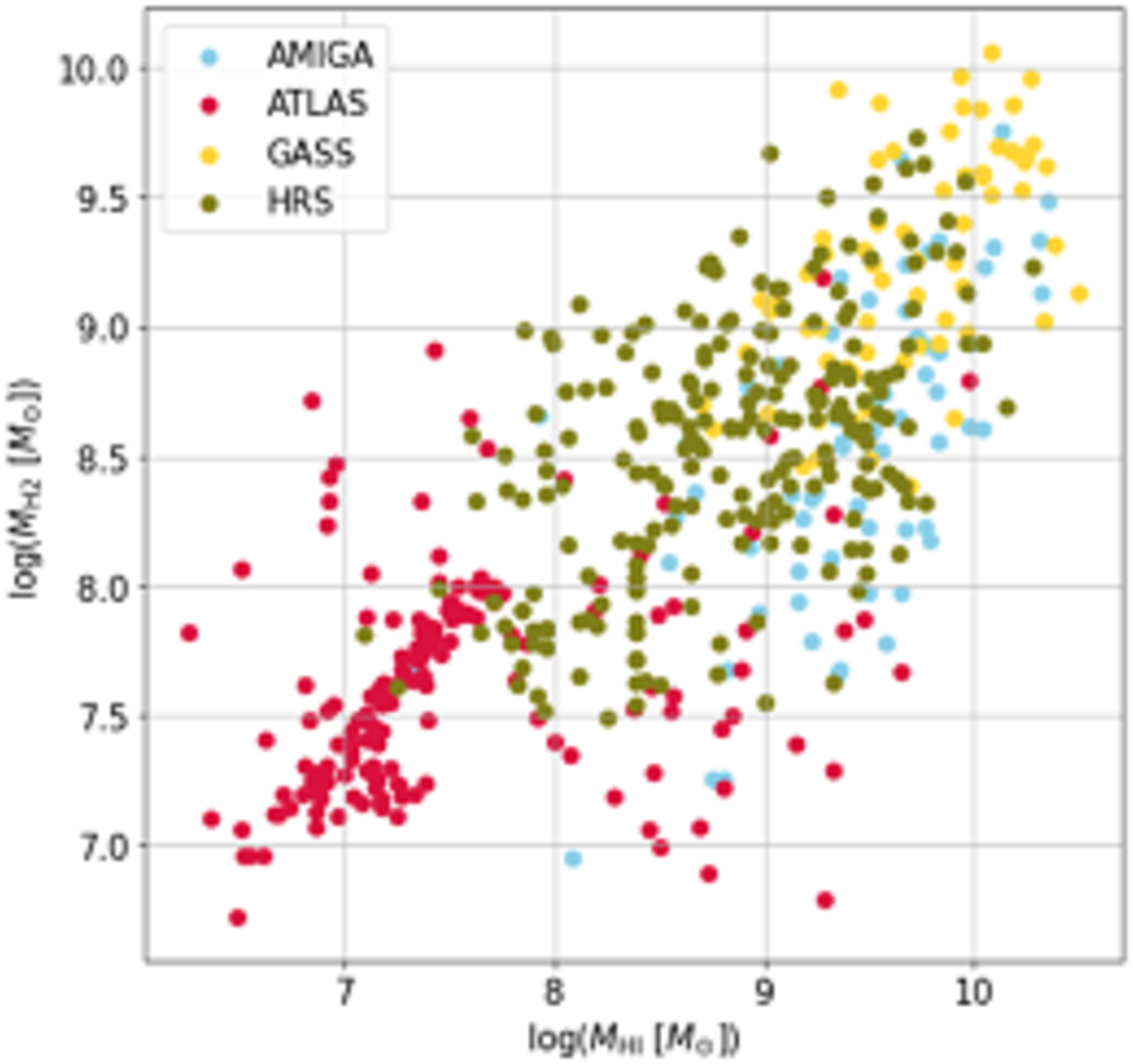}
\centering\includegraphics[width=0.47\textwidth]{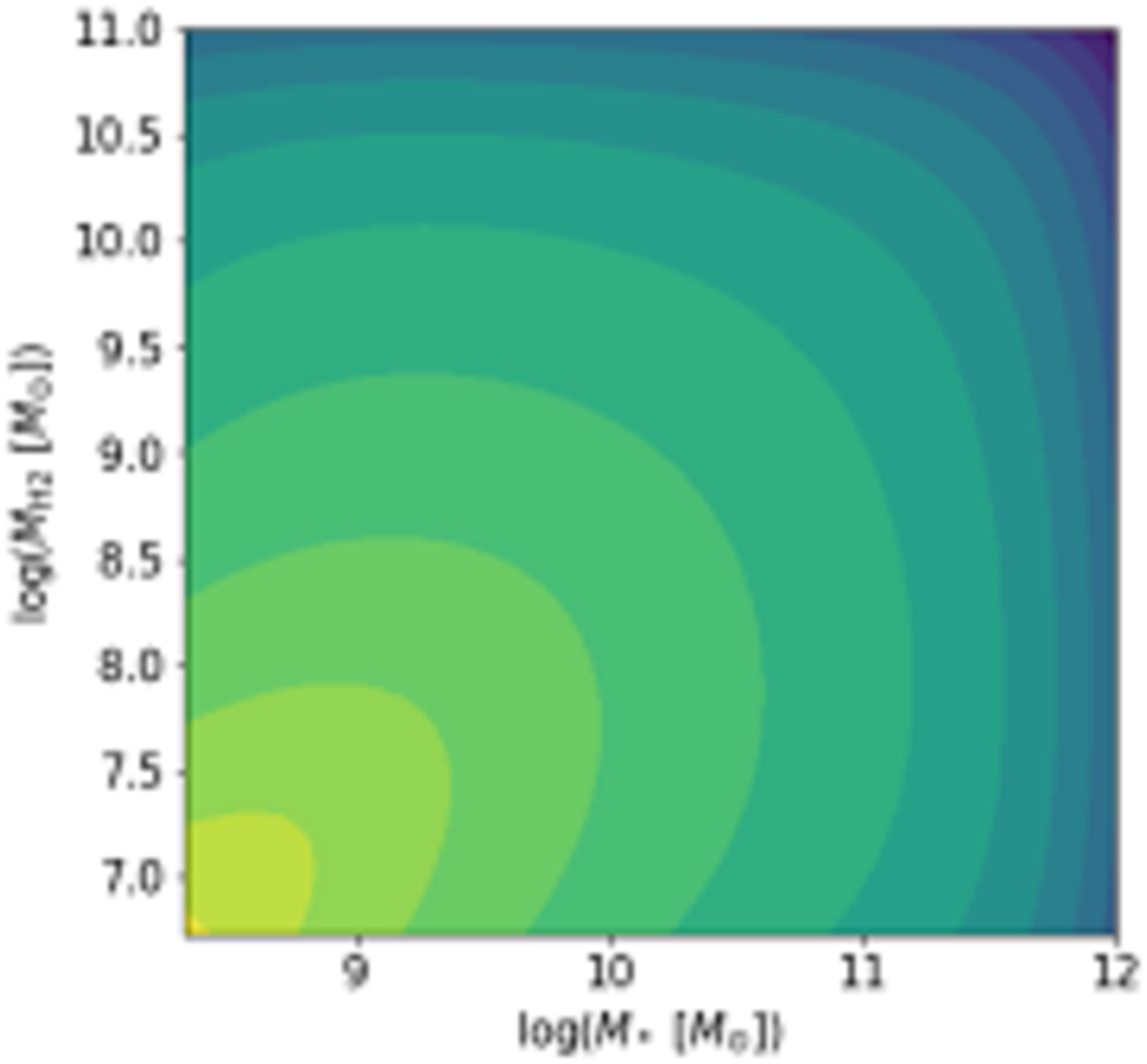}
\caption{Relation between $M_{\rm HI}$ and $M_{\rm H_2}$. 
Left: the data distribution on the $M_{\rm HI}$--$M_{\rm H_2}$ plane. 
Right: the corresponding bivariate PDF.  
The selected copula is the Frank copula with a parameter of $\delta = 8.80\pm0.43$. 
Data are taken from \citet{calette2018}. 
}
\label{fig:MHI_MH2}
\end{figure}

In contrast, the relation between $M_{\rm HI}$ and $M_{H_2}$ was found to be reproduced by the Frank copula, 
\begin{eqnarray}
  C(u_1, u_2;\delta) = -\frac{1}{\delta} \log \left[ 1- \frac{\left(e^{-\delta u_1} -1 \right)\left(e^{-\delta u_2}-1 \right)}{e^{\delta} - 1} \right], \quad (\delta \in \mathbb{R}\setminus \{0\});,
\end{eqnarray}
with a parameter $\delta = 8.80\pm0.43$.
Kendall's $\tau$ of this pair is $0.63$. 
The scatter plot of the data on the $M_{\rm HI}$--$M_{\rm H_2}$ plane, and the corresponding bivariate PDF are presented in Fig.~\ref{fig:MHI_MH2}. 
On this plane, We see a moderately strong dependence between these two variables, reflected to $\tau = 0.63$. 

For the conditional copula between $M_*$ and $M_{\rm HI}$ for given $M_{\rm H_2}$, the Student-$t$ copula
\begin{eqnarray}
    C(u_1,u_2;\rho,\nu)=\int^{u_1}_{-\infty}\int^{u_2}_{-\infty} \frac{1}{2\pi\sqrt{1-\rho^2}}{\left[1+\frac{s^2+t^2-2\rho st}{v(1-\rho^2)}\right]}^{-\frac{v+2}{2}} \pd s \pd t, \quad (\rho \in (-1,1),\ \nu>2);,
\end{eqnarray}
was selected as the appropriate copula with $\rho = 0.29\pm 0.03$ and $\nu=30.0$,respectively.  

\begin{figure}
\centering\includegraphics[width=0.43\textwidth]{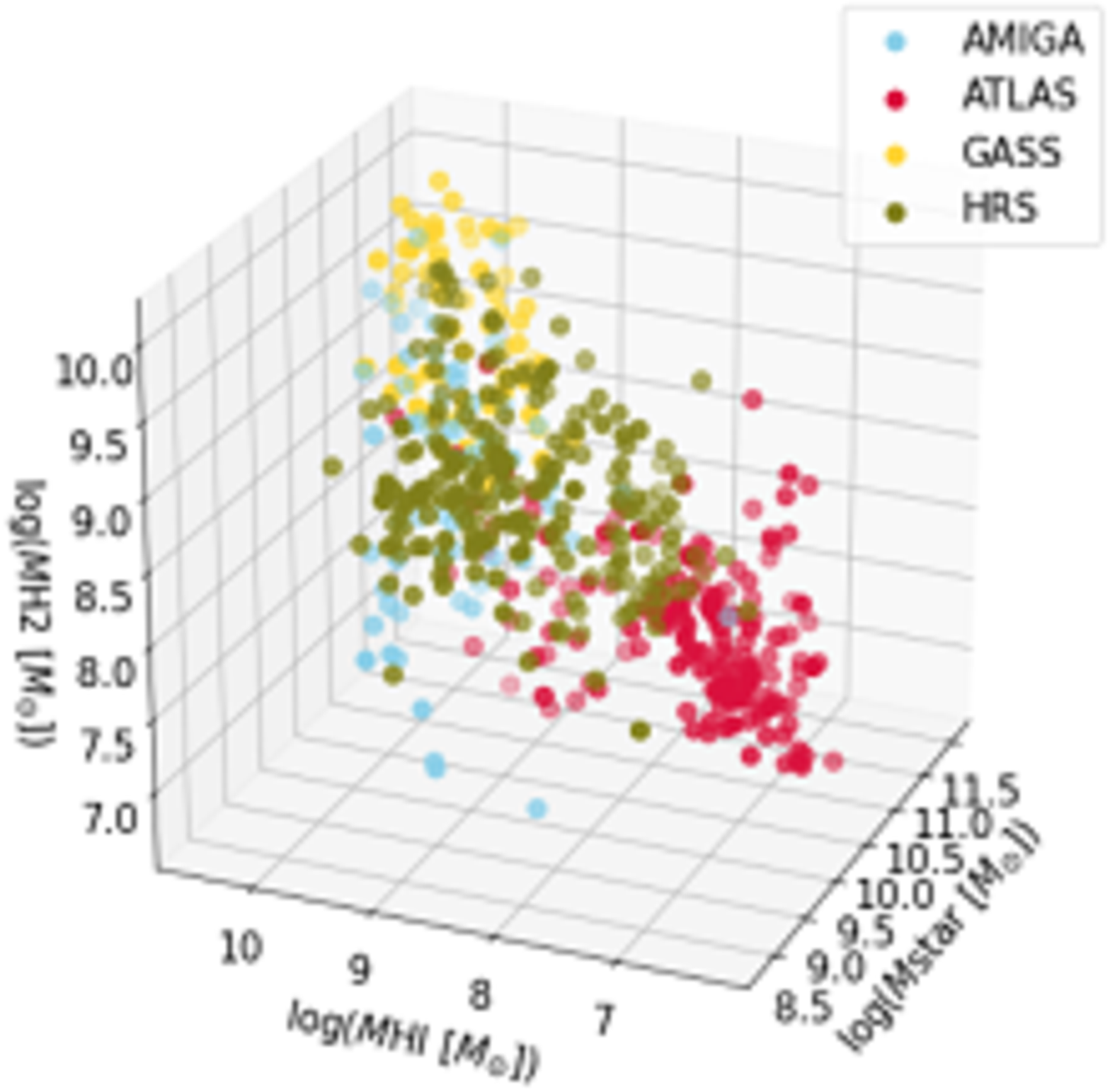}
\centering\includegraphics[width=0.47\textwidth]{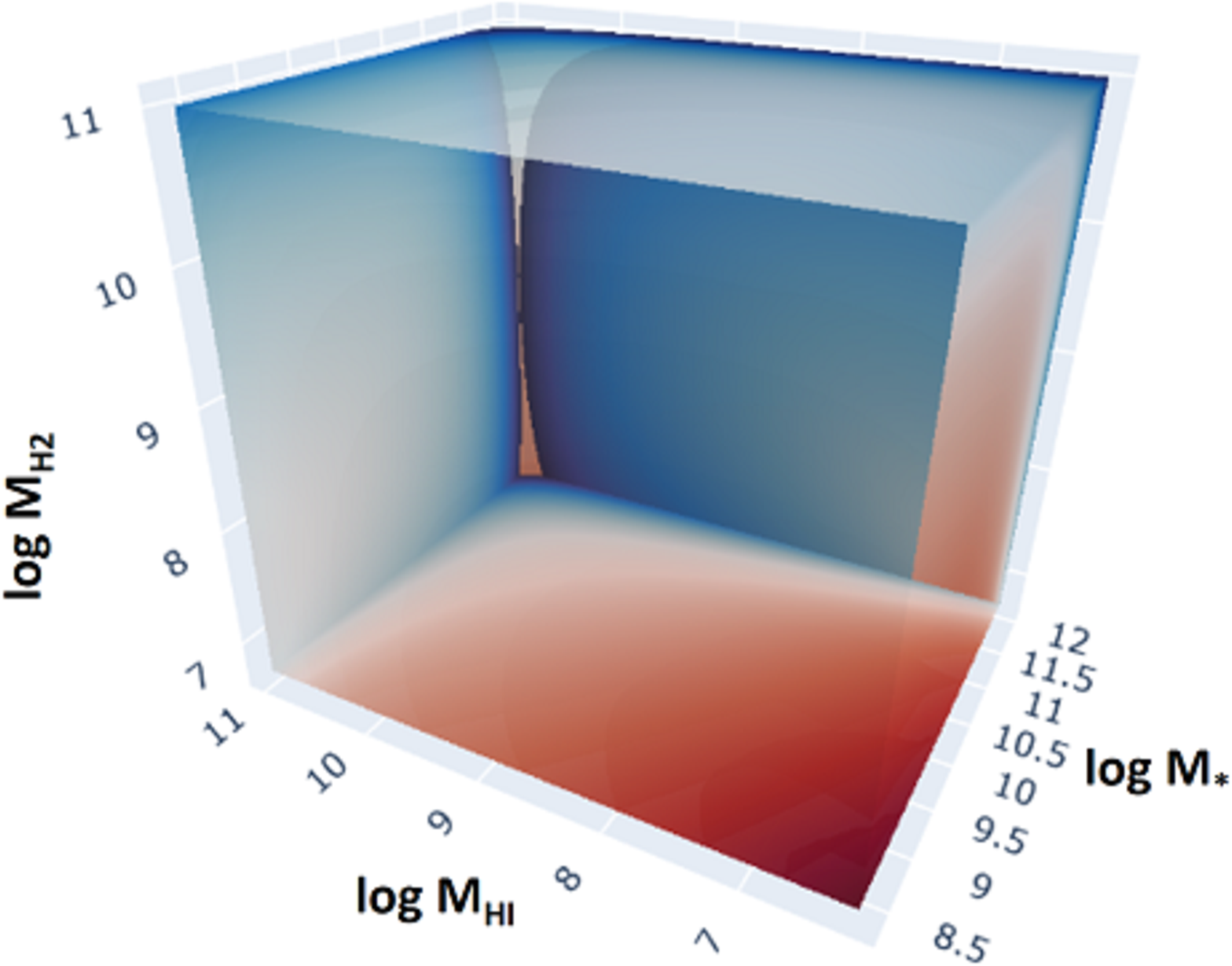}
\caption{The $M_*$-$M_{\rm HI}$-$M_{{\rm H}_2}$ 3-dim mass function obtained from the vine copula likelihood estimation. 
Left: the data distribution in the mass space. 
Right: the estimated 3-dim mass function (PDF). 
}
\label{fig:Mstar_MHI_MH2}
\end{figure}

Though it is not easy to visualize the 3-d structure of the PDF, its heavily asymmetric structure is well described in Fig.~\ref{fig:Mstar_MHI_MH2}. 
Of course it requires a further analysis for a physical interpretation, the copula method can provide us with a fundamentally important tool to understand the physical processes behind the multivariate LF/MF. 
The result here is just a demonstration of how the vine copula performs well for the multidimensional LF/MF estimation. 
More physical discussion is planned to be presented in our next work. 

If we go to much higher dimensions, human-intuitive approach may not work anymore even for a simple visualization. 
Very plausibly, we will confront the need for such a tremendously large data analysis. 
A machine-aided method will be a promising way to address such issues. 
We will discuss such a strategy in our future works.

\section{Summary and Conclusions}\label{sec:conclusion}

We have proposed a systematic method to build a bivariate luminosity or mass function of galaxies by using a copula (\citealt{takeuchi2010b}: T10). 
It allows us to construct a distribution function when only its marginal distributions are available and the dependence structure should be estimated from data. 

Though T10\nocite{takeuchi2010b} proposed a promising way to construct multivariate PDFs by a copula, the main limitation of the method is that it is not easy to extend a joint function to higher dimensions ($d > 2$), except some special cases like Gaussian. 
Even if we find such a multivariate analytic function in some special case, it would probably be very inflexible and impractical. 
In this work, we introduced a systematic method to extend the copula to unlimitedly higher dimensions by a pair-copula decomposition method, referred to as the vine copula. 

The vine copula method is extremely flexible because all the dependence structures are decomposed and reduced into pair dependence relations. 
We first formulated the factorization of a multidimensional DF/PDF.
This factorization is not unique, and we introduced a vine structure as a systematic method to sort out the complicated structure. 
The vine copula can be described intuitively by a diagrammatic method, as shown in Figs.~\ref{fig:vines_structure}, \ref{fig:d_vine}, and \ref{fig:c_vine}. 
We also presented that the likelihood parameter estimation of each copula and the model selection procedure can be performed simultaneously. 

Then, as an interesting and important example, we applied the vine copula PDF method to estimate the 3-d PDF of $M_*$, $M_{\rm HI}$, and $M_{\rm H_2}$. 
The vine copula can describe the PDF very well, and it can be used for the physical interpretation of the PDF as well as the evaluation of the complicated selection effect. 
We conclude that the vine copula method provides us with a promising way for the data analysis of unprecedentedly large surveys in the future. 
A machine-aided method will be a promising way to tackle such problems.

\section*{Acknowledgements}
First we thank  the referee, Alphonce Bere, for his careful reading of the manuscript and very useful suggestions. 
We are grateful to the members at the Institute for Statistical Mathematics, Shiro Ikeda, Satoshi Kuriki, Kenji Fukumizu, 
Yoh-ichi Mototake, Hideitsu Hino, and Mirai Tanaka for enlightening discussions.
We also thank Masami Ouchi, Kaiki Taro Inoue, Kiyoaki Christopher Omori, and Wen Shi for fruitful and useful comments. 
This work has been supported by JSPS Grants-in-Aid for Scientific Research (17H01110 and 19H05076). 
This work has also been supported in part by the Sumitomo Foundation Fiscal 2018 Grant for Basic Science Research Projects (180923), and the Collaboration Funding of the Institute of Statistical Mathematics ``New Development of the Studies on Galaxy Evolution with a Method of Data Science''.

\section*{Data availability statements}
The data underlying this article are available in the public domain mentioned in the main text.  
The datasets were derived from sources of this domain. 

%References


\begin{thebibliography}{} 
\bibitem[\protect\citeauthoryear{Aas et al.}{2009}]{aas2009}
  Aas K., Czado C., Frigessi A.\& Bakken H., 2009, Insurance: Mathematics and Economics, 44, 182

\bibitem[\protect\citeauthoryear{Aas}{2016}]{aas2016}
  Aas K. 2016, Econometrics, 4, 43.

\bibitem[\protect\citeauthoryear{Acar, Czado, Lysy}{Acar et al.}{2019}]{acar2019}
  Acar E.~F., Czado C., Lysy M., 2019, Econometrics and Statistics, 12, 181

\bibitem[\protect\citeauthoryear{Akaike}{1974}]{akaike1974}
  Akaike H.\ 1974, IEEE Trans. Autom. Contrib., 19, 716

\bibitem[\protect\citeauthoryear{Alidoost, Su, Stein}{Alidoost et al.}{2019}]{alidoost2019}
  Alidoost F., Su Z., Stein A., 2019, Weather and Climate Extremes, 26, 100227

\bibitem[\protect\citeauthoryear{Allen et al.}{2017}]{allen2017}
  Allen D.~E., McAleer M., Singh A.~K., 2017,  Sustainability, 9, 1762

\bibitem[\protect\citeauthoryear{Almeida, Chado, Manner}{Almeida et al.}{2016}]{almeida2016}
  Almeida C., Czado C., Manner H., 2016, Appl. Stochastic Models Bus. Ind. 2016, 32 62

\bibitem[\protect\citeauthoryear{Andreani, et al.}{2014}]{andreani2014}
  Andreani P., et al., 2014, A\&A, 566, A70

\bibitem[\protect\citeauthoryear{Andreani, et al.}{2018}]{andreani2018}
  Andreani P., Boselli A., Ciesla L., Vio R., Cortese L., Buat V., Miyamoto Y., 2018, A\&A, 617, A33

\bibitem[\protect\citeauthoryear{Bedford \& Cooke}{2001}]{bedford2001}
  Bedford, T.\ \& Cooke, R.~M., 2001, Annals of Mathematics and Artificial Intelligence, 32, 245

\bibitem[\protect\citeauthoryear{Bedford \& Cooke}{2002}]{bedford2002}
  Bedford, T., \& Cooke, R.~M., 2002, Ann.Stat., 30, 1031

\bibitem[\protect\citeauthoryear{Benabed et al.}{2009}]{benabed09}
  Benabed K., Cardoso J.-F., Prunet S., Hivon E., 2009, MNRAS, 400, 219 
  
\bibitem[\protect\citeauthoryear{Bhatawdekar, et al.}{2019}]{bhatawdekar2019}
  Bhatawdekar R., Conselice C.~J., Margalef-Bentabol B., Duncan K., 2019, MNRAS, 486, 3805

\bibitem[\protect\citeauthoryear{Binggeli, Sandage, \& Tammann}{1988}]{binggeli88}
  Binggeli B., Sandage A., Tammann G.~A., 1988, ARA\&A, 26, 509 

\bibitem[\protect\citeauthoryear{Blanton et al.}{2001}]{blanton01}
  Blanton M.~R., et al., 2001, AJ, 121, 2358 

%\bibitem[\protect\citeauthoryear{Boselli et al.}{2001}]{boselli01}
%  Boselli A., Gavazzi G., Donas J., Scodeggio M., 2001, AJ, 121, 753 

\bibitem[\protect\citeauthoryear{Boselli, et al.}{2010}]{boselli2010}
  Boselli A., et al., 2010, PASP, 122, 261

\bibitem[\protect\citeauthoryear{Boselli, Cortese \& Boquien}{2014}]{boselli2014a} 
  Boselli A., Cortese L., Boquien M., 2014, A\&A, 564, A65

\bibitem[\protect\citeauthoryear{Boselli, et al.}{2014b}]{boselli2014b} 
  Boselli A., Cortese L., Boquien M., Boissier S., Catinella B., Lagos C., Saintonge A., 2014, A\&A, 564, A66

\bibitem[\protect\citeauthoryear{Boselli, et al.}{2014c}]{boselli2014c} 
  Boselli A., et al., 2014, A\&A, 564, A67

\bibitem[\protect\citeauthoryear{Bradford, Geha \& Blanton}{2015}]{bradford2015} 
  Bradford J.~D., Geha M.~C., Blanton M.~R., 2015, ApJ, 809, 146

\bibitem[\protect\citeauthoryear{Brechmann \& Schepsmeier}{2013}]{brechmann2013}
  Brechmann, E.~C. \& Schepsmeier, U., 2013, Journal of Statistical Software, 52, 1

\bibitem[\protect\citeauthoryear{Callau Poduje \& Haberlandt}{2018}]{callau_poduje2018} 
  Callau Poduje A.~C., Haberlandt U., 2018, Water, 10, 862

\bibitem[\protect\citeauthoryear{Calette, et al.}{2018}]{calette2018} 
  Calette A.~R., Avila-Reese V., Rodr{\'\i}guez-Puebla A., Hern{\'a}ndez-Toledo H., Papastergis E., 2018, RMxAA, 54, 443

\bibitem[\protect\citeauthoryear{Caplar, Lilly \& Trakhtenbrot}{2018}]{caplar2018} 
  Caplar N., Lilly S.~J., Trakhtenbrot B., 2018, ApJ, 867, 148

\bibitem[\protect\citeauthoryear{Catinella, et al.}{2013}]{catinella2013} 
  Catinella B., et al., 2013, MNRAS, 436, 34

\bibitem[\protect\citeauthoryear{Chapman et al.}{2003}]{chapman03}
  Chapman S.~C., Helou G., Lewis G.~F., Dale D.~A., 2003, ApJ, 588, 186 
  
\bibitem[\protect\citeauthoryear{Cho{\l}oniewski}{1985}]{choloniewski85} 
  Cho{\l}oniewski J., 1985, MNRAS, 214, 197 

\bibitem[\protect\citeauthoryear{de Lapparent et al.}{2003}]{delapparent03}
  de Lapparent V., Galaz G., Bardelli S., Arnouts S., 2003, A\&A, 404, 831 

\bibitem[\protect\citeauthoryear{D'Souza, Vegetti \& Kauffmann}{2015}]{dsouza2015}
  D'Souza R., Vegetti S., Kauffmann G., 2015, MNRAS, 454, 4027

\bibitem[\protect\citeauthoryear{Dutta, Khandai \& Dey}{2020}]{dutta2020}
  Dutta S., Khandai N., Dey B., 2020, MNRAS, 494, 2664

\bibitem[\protect\citeauthoryear{Gr\"{a}ler}{2011}]{graler2011}
  Gr\"{a}ler B., Pebesma E., 2011, Procedia Environmental Sciences, 7,206

\bibitem[\protect\citeauthoryear{Gr\"{a}ler}{2014}]{graler2014}
  Gr\"{a}ler B., 2014, Spatial Statistics, 10, 102

\bibitem[\protect\citeauthoryear{Gunawardhana, et al.}{2015}]{gunawardhana2015}
  Gunawardhana M.~L.~P., et al., 2015, MNRAS, 447, 875

\bibitem[\protect\citeauthoryear{Hern{\'a}ndez-Toledo, et al.}{2010}]{hernandez_toledo2010} 
  Hern{\'a}ndez-Toledo H.~M., V{\'a}zquez-Mata J.~A., Mart{\'\i}nez-V{\'a}zquez L.~A., Choi Y.-Y., Park C., 2010, AJ, 139, 2525

\bibitem[\protect\citeauthoryear{J\"{a}ger \& al.}{2017}]{jager2017} 
 J\"{a}ger, Morales N\'{a}poles O., 2017, ASCE-ASME J. Risk Uncertainty Eng. Syst., Part A: Civ. Eng., 3, 04017014 

\bibitem[\protect\citeauthoryear{Jiang et al.}{2009}]{jiang09} 
  Jiang I.-G., Yeh L.-C., Chang Y.-C., Hung W.-L., 2009, AJ, 137, 329 

\bibitem[\protect\citeauthoryear{Jiang, Yeh \& Hung}{2015}]{jiang2015}
  Jiang I.-G., Yeh L.-C., Hung W.-L., 2015, MNRAS, 449, L65

\bibitem[\protect\citeauthoryear{Jo}{2019}]{jo2019a}
  Jo H.-H., 2019, PhRvE, 100, 012306

\bibitem[\protect\citeauthoryear{Jo, et al.}{2019}]{jo2019b}
  Jo H.-H., Lee B.-H., Hiraoka T., Jung W.-S., 2019, PhRvE, 100, 022307
  
\bibitem[\protect\citeauthoryear{Johnson \& Kotz}{1977}]{johnson77}
  Johnson N.\ L., Kotz S., 1977, Comm.\ Statist.\ Ser.\ A (Theory and Methods),
  6, 485
\bibitem[\protect\citeauthoryear{Johnston}{2011}]{johnston2011}
  Johnston R., 2011, A\&ARv, 19, 41

\bibitem[\protect\citeauthoryear{Johnston, Teodoro \& Hendry}{2012}]{johnston2012}
  Johnston R., Teodoro L., Hendry M., 2012, MNRAS, 421, 270

\bibitem[\protect\citeauthoryear{Jones, et al.}{2018}]{jones2018}
  Jones M.~G., Haynes M.~P., Giovanelli R., Moorman C., 2018, MNRAS, 477, 2

\bibitem[\protect\citeauthoryear{Keres, Yun \& Young}{2003}]{keres2003}
  Keres D., Yun M.~S., Young J.~S., 2003, ApJ, 582, 659

\bibitem[\protect\citeauthoryear{Khuntia et al. }{2019}]{khuntia2019}
  Khuntia S.~R., Rueda, J.~L., van der Meijden M.~A.~M.~M., 2019, Wind Energy, 1

\bibitem[\protect\citeauthoryear{Koen}{2009}]{koen09}
  Koen C., 2009, MNRAS, 393, 1370 

\bibitem[\protect\citeauthoryear{Koen \& Bere}{2017}]{koen2017}
  Koen C., Bere A., 2017, MNRAS, 471, 2771

\bibitem[\protect\citeauthoryear{Koprowski, et al.}{2017}]{koprowski2017}
  Koprowski M.~P., et al., 2017, MNRAS, 471, 4155

\bibitem[\protect\citeauthoryear{Kotz, Balakrishnan, \& Johnson}{2000}]{kotz00}
  Kotz S., Balakrishnan N., Johnson N., L., 2000, Continuous Multivariate
  Distributions, Volume 1: Models and Applications, 2nd ed., John Wiley \& Sons,
  New York, pp.51--62

\bibitem[\protect\citeauthoryear{Kloubert}{2020}]{kloubert2020} 
  Kloubert M.-L., 2020, Energies, 13, 1727

\bibitem[\protect\citeauthoryear{Kurowicka \& Joe}{2011}]{kurowicka2011}
  Kurowicka D., Joe H., 2011, Dependence Modeling: a Vine Copula Handbook, World Scientific

\bibitem[\protect\citeauthoryear{Lin \& Kilbinger}{2015}]{lin2015}
  Lin C.-A., Kilbinger M., 2015, A\&A, 583, A70

\bibitem[\protect\citeauthoryear{Lin et al.}{2014}]{lin2014}
  Lin G.~D., Dou X., Kuriki S. et al., 2014, J.~Stat Distrib~App, 1, 14

\bibitem[\protect\citeauthoryear{Mejdoub \& Ben Arab}{2018}]{mejdoub2018}
  Mejdoub H., Ben Arab M., 2018, Research in International Business and Finance, 45(C), 208

\bibitem[\protect\citeauthoryear{Lake, et al.}{2017}]{lake2017} 
  Lake S.~E., Wright E.~L., Tsai C.-W., Lam A., 2017, AJ, 153, 189

\bibitem[\protect\citeauthoryear{Lisenfeld, et al.}{2011}]{lisenfeld2011} 
  Lisenfeld U., et al., 2011, A\&A, 534, A102

\bibitem[\protect\citeauthoryear{L{\'o}pez-Sanjuan, et al.}{2017}]{lopez_sanjuan2017}
  L{\'o}pez-Sanjuan C., et al., 2017, A\&A, 599, A62

\bibitem[\protect\citeauthoryear{Mashian, Oesch \& Loeb}{2016}]{mashian2016}
  Mashian N., Oesch P.~A., Loeb A., 2016, MNRAS, 455, 2101

\bibitem[\protect\citeauthoryear{Mobasher, Sharples, \& Ellis}{Mobasher et al.}{1993}]{mobasher93}
  Mobasher B., Sharples R.\ M., Ellis R.\ S., 1993, MNRAS, 263, 560

\bibitem[\protect\citeauthoryear{Moffett, et al.}{2016}]{moffett2016} 
  Moffett A.~J., et al., 2016, MNRAS, 462, 4336

\bibitem[\protect\citeauthoryear{Nagler et al.}{2019}]{nagler2019}
  Nagler T.,Bumann C., Czado C., 2019, Journal of Multivariate Analysis, 172, 180

\bibitem[\protect\citeauthoryear{Nelsen}{2006}]{nelsen06}
  Nelsen R. B., 2006, An Introduction to copulae, 2nd ed., Springer, New York, \S 2

\bibitem[\protect\citeauthoryear{Peters et al.}{2014}]{peters2014}
  Peters G.~W., Dong A.~X.~D., Kohn R., 2014, Insurance: Mathematics and Economics, 59, 258

\bibitem[\protect\citeauthoryear{Rodr{\'\i}guez-Puebla, et al.}{2020}]{rodriguez_puebla2020} 
  Rodr{\'\i}guez-Puebla A., Calette A.~R., Avila-Reese V., Rodriguez-Gomez V., Huertas-Company M., 2020, PASA, 37, e024

\bibitem[\protect\citeauthoryear{Saintonge, et al.}{2011}]{saintonge2011} 
  Saintonge A., et al., 2011, MNRAS, 415, 32

\bibitem[\protect\citeauthoryear{Sato, Ichiki \& Takeuchi}{2010}]{sato2010}
  Sato M., Ichiki K., Takeuchi T.~T., 2010, PhRvL, 105, 251301

\bibitem[\protect\citeauthoryear{Sato, Ichiki \& Takeuchi}{2011}]{sato2011}
  Sato M., Ichiki K., Takeuchi T.~T., 2011, PhRvD, 83, 023501

\bibitem[\protect\citeauthoryear{Saunders et al.}{2000}]{saunders00}
  Saunders W., et al., 2000, MNRAS, 317, 55 
 
\bibitem[\protect\citeauthoryear{Schechter}{1976}]{schechter76}
  Schechter P.\ L., 1976, ApJ, 203, 297

\bibitem[\protect\citeauthoryear{Schafer}{2007}]{schafer07}
  Schafer C.\ M., 2007, ApJ, 661, 703

\bibitem[\protect\citeauthoryear{Scherrer et al.}{2010}]{scherrer10}
  Scherrer R.~J., Berlind A.~A., Mao Q., McBride C.~K., 2010, ApJ, 708, L9 

\bibitem[\protect\citeauthoryear{Schwarz}{1978}]{schwarz1978}
  Schwarz G., 1978, Ann.~Stat., 6, 461

\bibitem[\protect\citeauthoryear{Serra, et al.}{2012}]{serra2012}
  Serra P., et al., 2012, MNRAS, 422, 1835

\bibitem[\protect\citeauthoryear{Shi \& Yang}{2018}]{shi2018}   Shi P., Yang L., 2018, Journal of the American Statistical Association, 113, 122

\bibitem[\protect\citeauthoryear{Simon \& Schneider}{2017}]{simon2017}
  Simon P., Schneider P., 2017, A\&A, 604, A109

\bibitem[\protect\citeauthoryear{Sklar}{1959}]{sklar59}
  Sklar A., 1959, Publ. Inst. Stat. Univ. Paris, 8, 229

\bibitem[\protect\citeauthoryear{Sriboonchitta et al.}{2014}]{sriboonchitta2014}
  Sriboonchitta S., Liu J., Kreinovich V., Nguyen H.T., 2014 in  Modeling Dependence in Econometrics. Advances in Intelligent Systems and Computing, vol 251, Huynh VN., Kreinovich V., Sriboonchitta S. (eds), Springer, Cham

\bibitem[\protect\citeauthoryear{Takeuchi}{2000}]{takeuchi00a}
  Takeuchi T.\ T., 2000, Ap\&SS, 271, 213

\bibitem[\protect\citeauthoryear{Takeuchi, Yoshikawa, \& Ishii}
{Takeuchi et al.}{2000}]{takeuchi00b}
  Takeuchi T.\ T., Yoshikawa K., Ishii T.\ T., 2000, ApJS, 129, 1

\bibitem[\protect\citeauthoryear{Takeuchi, Yoshikawa, \& Ishii}
{Takeuchi et al.}{2003b}]{takeuchi03b}
  Takeuchi T.\ T.,  Yoshikawa K., Ishii T.\ T., 2003, ApJ, 587, L89
 
\bibitem[\protect\citeauthoryear{Takeuchi, Buat, \& Burgarella}{Takeuchi et al.}{2005c}]{takeuchi05c}
  Takeuchi T.~T., Buat V., Burgarella D., 2005, A\&A, 440, L17 
 
\bibitem[\protect\citeauthoryear{Takeuchi et al.}{2010a}]{takeuchi10a}
  Takeuchi T.~T., Buat V., Heinis S., Giovannoli E., Yuan F.~-T., Iglesias-Paramo J., Murata K.~L., Burgarella D., 2010, A\&A, 514, A4

\bibitem[\protect\citeauthoryear{Takeuchi}{2010}]{takeuchi2010b}
  Takeuchi T.~T., 2010, MNRAS, 406, 1830 (T10)

\bibitem[\protect\citeauthoryear{Takeuchi, et al.}{2013}]{takeuchi2013}
  Takeuchi T.~T., Sakurai A., Yuan F.-T., Buat V., Burgarella D., 2013, EP\&S, 65, 281
  
\bibitem[\protect\citeauthoryear{Torabi et al.}{2020}]{torabi2020}
  Torabi S., Dourandish A., Daneshvar Kakhki M., Kianirad A., Mohammadi H., 2020,  in Rashidghalam M. (eds) The Economics of Agriculture and Natural Resources. Perspectives on Development in the Middle East and North Africa (MENA) Region, Springer, Singapore

\bibitem[\protect\citeauthoryear{Trivedi \& Zimmer}{2005}]{trivedi05}
  Trivedi P.\ R., Zimmer D.\ M., 2005, Foundations and Trends in Econometrics, 1, 1

\bibitem[\protect\citeauthoryear{Vallini, et al.}{2016}]{vallini2016} 
  Vallini L., Gruppioni C., Pozzi F., Vignali C., Zamorani G., 2016, MNRAS, 456, L40

\bibitem[\protect\citeauthoryear{Vio, Nagler, Andreani}{Vio et al.}{2020}]{vio2020}
  Vio R., Nagler T.~W., Andreani P., 2020, arXiv, arXiv:2006.06268

\bibitem[\protect\citeauthoryear{Willmer et al.}{2006}]{willmer06}
  Willmer C.~N.~A., et al., 2006, ApJ, 647, 853 

\bibitem[\protect\citeauthoryear{Wright, et al.}{2017}]{wright2017} 
  Wright A.~H., et al., 2017, MNRAS, 470, 283

\bibitem[\protect\citeauthoryear{Xu D.\ et al.}{2017}]{xu_d2017}
  Xu D., Wei Q., Elsayed E.~A., Chen Y., Kang R., 2017, Qual. Reliab. Engng. Int., 33 803

\bibitem[\protect\citeauthoryear{Xu D. et al.}{2018}]{xu_d2018}
  Xu D., He J., Sui S., Jiang S., Zhang W., 2018, IEEE Access, 6, 3120 

\bibitem[\protect\citeauthoryear{Xu, Hua, Xu}{Xu M.\ et al.}{2017}]{xu_m2017}
 Xu M., Hua L., Xu S., 2017, Technometrics, 59, 508


\bibitem[\protect\citeauthoryear{Young, et al.}{2011}]{young2011} 
  Young L.~M., et al., 2011, MNRAS, 414, 940

\bibitem[\protect\citeauthoryear{Yuan \& Wang}{2013}]{yuan2013}
  Yuan Z., Wang J., 2013, Ap\&SS, 345, 305

\bibitem[\protect\citeauthoryear{Yuan, et al.}{2018}]{yuan2018}
  Yuan Z., Wang J., Worrall D.~M., Zhang B.-B., Mao J., 2018, ApJS, 239, 33

\end{thebibliography}
\end{document}